\documentclass[letterpaper]{JHEP3}

\usepackage{graphicx}

\def\Box{{\hbox{$\sqcup$}\llap{\hbox{$\sqcap$}}}}

\def \lsim{\mathrel{\vcenter
     {\hbox{$<$}\nointerlineskip\hbox{$\sim$}}}}
\def \gsim{\mathrel{\vcenter
     {\hbox{$>$}\nointerlineskip\hbox{$\sim$}}}}

\def\bea{\begin{eqnarray}}
\def\eea{\end{eqnarray}}
\def\nn{\nonumber}

\def\exd{{\rm d}}
\def\pref#1{(\ref{#1})}
\def\endignore{}
\def\ignore #1\endignore{} 

\def\bd{\begin{displaymath}}
\def\ed{\end{diplaymath}}

\def\KK{{\scriptscriptstyle KK}}
\def\ssM{{\scriptscriptstyle M}}

\def\ta{{A}}

\def\cA{{\cal A}}

\def\cF{{\cal F}}
\def\cK{{\cal K}}

\def\ba{\begin{eqnarray}}
\def\ea{\end{eqnarray}}
\def\be{\begin{equation}}
\def\ee{\end{equation}}

\def\exd{{\rm d}}

\newcommand{\comment}[1]{}
\newcommand{\tg}{\tilde{g}}

\newcommand{\beq}{\begin{equation} }
\newcommand{\eeq}{\end{equation}}

\newcommand{\lb}{\label}

\newcommand{\im}{\textnormal{Im}}

\newcommand{\roughly}[1]{\raise.3ex\hbox{$#1$\kern-.75em\lower1ex\hbox{$\sim$}}}

\title{Warped
Supersymmetry Breaking}

\author{C.P. Burgess,$^1$ P.G. C\'amara,$^2$  S.P. de Alwis,$^{3}$
S.B. Giddings,$^{4,5}$  A. Maharana,$^4$ \qquad\qquad\qquad
F. Quevedo,$^{5,6}$  K. Suruliz$^6$
\\

$^1$Department of Physics and Astronomy, McMaster University, Hamilton ON, Canada\\
\qquad and Perimeter Institute for Theoretical Physics, Waterloo ON, Canada.\\

$^2$ CPHT, UMR du CNRS 7644, Ecole Polytechnique, 91128 Palaiseau, France.\\

$^3$ Physics Department, University of Colorado, Boulder CO 80309, USA.\\

$^4$ Department of Physics, University of California, Santa Barbara CA 93106-9530, USA.\\

$^5$ KITP, University of California, Santa Barbara CA 93106-4030, USA.\\

$^6$ DAMTP/CMS, University of Cambridge, Wilberforce Road, Cambridge CB3 0WA, UK.\\  }
\date{}

\vspace{-0.2cm}\abstract {
We address the size of supersymmetry-breaking effects
within
string theory
settings where the observable sector
resides deep within a strongly warped region, with supersymmetry
breaking not necessarily localized in that region. Our particular
interest is in how the supersymmetry-breaking scale seen by the
observable sector depends on this warping.
We focus concretely on type IIB flux compactifications and
obtain this dependence in two ways: by computing within the microscopic
string theory supersymmetry-breaking masses in
Dp-brane
supermultiplets; and by investigating how warping gets encoded
into masses within the low-energy 4D effective theory.
We identify two different ways to identify `the' 4D gravitino in
such systems -- the state whose supersymmetry is the least broken,
and the state whose couplings are the most similar to the 4D
graviton's -- and argue that these need not select the same state
in strongly warped settings.
We formulate the conditions required for the existence of a
description in terms of a 4D SUGRA formulation, or in terms of 4D
SUGRA together with soft-breaking terms, and describe in
particular situations where neither exist for some
non-supersymmetric compactifications. We suggest that some effects
of warping are captured by a linear $A$ dependence in the K\"ahler
potential. We outline some implications of our results for the
KKLT scenario of moduli stabilization with broken SUSY.}

\preprint{NSF-KITP-06-87\\ DAMTP-2006-78 \\ CPHT-RR069.0906\\ hep-th/0610255}

\begin{document}

\section{Introduction}

Understanding how supersymmetry breaks has been a Holy Grail for
string theorists for decades, because it is likely a crucial
prerequisite for understanding string theory's low-energy
predictions. Like the search for the Grail this has proven to be
an elusive quest, complicated as it is by related issues of
modulus stabilization. Considerable progress has come recently,
however, with the recognition that Type IIB string vacua can
stabilize many moduli in the presence of fluxes \cite{sethi1,GKP}.

The remarkable properties of the warped compactifications to which
these studies lead open up potentially interesting new
possibilities for constructing phenomenologically attractive
string vacua, both for applications to particle physics
\cite{PPApps} and to cosmology \cite{CosmoApps}. They do so for
several reasons. First, by providing a plausible setting in
which all moduli may be fixed \cite{GKP,KKLT}
they provide a concrete laboratory within which to compute how
supersymmetry breaks. This potentially represents a great leap
forward since it allows one to deal with a serious drawback of
previous calculations of supersymmetry-breaking effects for string
vacua (for a review see \cite{OldSoftSUSY}). Because these earlier
calculations did not construct the potential which stabilized the
various moduli, they could not determine the values of the moduli.
In particular, they could not address whether or not supersymmetry
was restored at a minimum of the moduli potential, as is very
often the case in practice.

The second reason these warped compactifications have been so
intriguing for phenomenology is that they can contain
strongly-warped regions (`throats') within which the energies of
localized states can be strongly suppressed by gravitational
redshift. This possibility is very interesting since the energy
gained by localization can dominate the energy cost due to the
gradients which the localization requires. This can dramatically
change the kinds of states which dominate the low-energy world,
with localized states often being preferred over those which
spread to fill all of the compact dimensions. Among the
consequences of this observation is the possibility of having new
ways to obtain large hierarchies of scale, such as by having all
Standard Model degrees of freedom localized in a region of strong
warping \cite{RSone,RStwo,HVer,KlSt}. It also opens up new ways
for energy to be efficiently channelled into such throats,
potentially providing new ways to think about reheating during
early-universe cosmology \cite{BBC,WarpedReheat2}.

%


Our goal in this paper is to analyze the size of
supersymmetry-breaking effects within highly-warped
compactifications of string theory, with emphasis on when the
low-energy physics can be captured by an effective 4D description,
and on how this description depends on the underlying scales. We
focus on supersymmetry breaking triggered by nonzero (0,3) bulk
fluxes within Type IIB vacua (for related work in this direction
see \cite{kim}), and where necessary we imagine matter fields are
identified with open-string degrees of freedom in the world-volume
of D-branes, being suitably described at low energies by the
corresponding Dirac-Born-Infeld and Chern-Simons actions. Within
this framework we follow how supersymmetry breaking depends on
three important parameters: $\vartheta$ the strength of
supersymmetry breaking fluxes \footnote{$\vartheta$
will correspond to $W_{0}$ in the regime
that a 4D effective description is valid.};  $e^{A_m}$ the minimal value of
warping, as well as the compactification volume, ${\cal V} = V
/ {\alpha'}^3$.



While reproducing standard results in the limit of small warping,
we find several interesting differences when the warping is large.
In particular, two natural criteria for identifying `the'
gravitino of the low-energy 4D theory point to different states.
The first of these, the criterion of the least broken
supersymmetry, selects the lightest of all gravitino Kaluza-Klein
(KK) modes (or equivalently, the gravitino state which becomes
massless if supersymmetry breaking is adiabatically switched off).
In the presence of strong warping we argue that the wave-function
for this state is localized within the strongly warped region, as
is generically true for KK modes
\cite{Goldberger,hepth0507158,hepth0603233}. However, because of
this strong warping the strength of the interactions of this state
are generically set by a warped mass scale, which can be much
smaller than the 4D Planck mass, $M_p$. This is what makes this
state differ in general with the state selected by the second
criterion: that whose interactions are most similar to the
massless 4D graviton.

What we find echoes what is known about supersymmetry breaking in
Randall-Sundrum geometries, as has been extensively studied in the
context of $S^1/\mathbb{Z}_2$ orbifold compactifications of 5D
gauged supergravity on AdS backgrounds
\cite{SusyRS1,SusyRS2,GhPom}. In these setups, the observable
(Standard Model) sector generically resides on the infrared brane,
at the strongly warped end of the orbifold, and supersymmetry
breaking is mediated by the radial modulus and/or the Weyl
anomaly. Both of the gravitino states described above have a
simple physical interpretation in these models when the dual CFT
description of the throat is used, with the lightest gravitino
describing a spin-3/2 resonance of the strongly interacting CFT
dynamics (which gauges an emergent supersymmetry, not simply
related to the supersymmetry of the constituent states). In this
instance it is the second criterion which gives the gravitino
which partners the graviton in its couplings to the `constituent'
CFT degrees of freedom.

Although we argue that generically a 4D description need not be
possible for strongly warped systems, when it is we find that
strong warping alters the resulting low-energy K\"ahler potential
for Type IIB string compactifications. We identify a universal
contribution of this kind, which leads to exponential suppressions
for F-terms in the low-energy theory. This is similar to what is
seen for supersymmetric RS systems \cite{KWarp}, where more
explicit forms for the warping-induced modifications to the
K\"ahler potential are possible because of the simplicity of the
underlying 5D AdS geometry.


Our presentation comes in two parts. In the next section,
\ref{warpsec2}, we provide a generic semi-quantitative discussion
of the scales which arise in supersymmetry-breaking
compactifications with warped extra dimensions. (Although some
estimates and explicit calculations of supersymmetry breaking
scales have been made elsewhere \cite{hepth0208123}, these focus
on only one of the two types of 4D gravitino mentioned above.)
This section also gives insights on how the lightest gravitino
mass undergoes warping suppression, illustrating the phenomenon
with the aid of a scalar field toy model. We provide here a
criterion for when to expect the low-energy limit to be described
by a 4D effective theory, and when this effective theory should be
described by a 4D supergravity, by a supergravity supplemented by
soft-breaking terms, or by a generic non-supersymmetric field
theory. Section \ref{warpsec3} then follows with simple calculable
examples of flux-induced mass terms for D3 and D7 branes, together
with a discussion of an approach to summarizing the warping
dependence of the mass in the low-energy 4D effective field
theory. We describe aspects of phenomenology in \S \ref{warpsec4},
and then close in \S \ref{warpsec5} with a short summary and some
concluding remarks.

\section{Mass scales and effective descriptions}
\label{warpsec2} We now turn to a qualitative description of the
scales which arise when higher-dimensional supergravities are
compactified on strongly warped geometries.\footnote{For some
earlier general discussion of scales see  \cite{hepth0208123}. For
some discussion of the problems that arise in trying to derive an
N=1 SUGRA potential in the context of non-trivial warping see
\cite{GhPom}, \cite{deAlwis:2003sn} and \cite{KKA}; for discussion
of a prescription to derive the potential and other aspects of the
effective theory see \cite{hepth0507158}. } With a view to
phenomenological applications, our interest is in particular on
the scales and the effective field theory which are relevant to
the 4D dynamics of the low-energy theory (possibly associated with
particles which are localized deep within a strongly warped
region), with supersymmetry broken in some other sector due to
fluxes or some other supersymmetry-breaking effects.

We start with a general description of the scales and effective
theories which can arise within strongly warped compactifications,
before returning later to describe their relevance for the scale
of supersymmetry breaking as seen by observers localized in the
warped regions.

\subsection{Mass scales in warped throats}

Since many of the issues which arise for supersymmetry breaking in
warped environments are generic to the kinematics of warping, we
start here by summarizing some general properties of warped
compactifications, together with examples from IIB compactifications.

\subsubsection{KK excitations and scales}
\label{warp211}

It is useful to begin with a reminder of the mass scales which
arise within unwarped string compactifications, having a
product-space vacuum configuration
\be
    \exd s^2 = \eta_{\mu\nu} \exd x^\mu \exd x^\nu + g_{mn}(y) \exd y^m
    \exd y^n \,,
\ee
together with configurations for the other bosonic supergravity
fields. Our conventions are to use 10D indices $M,N,\ldots
=0,\ldots,9$ for spacetime; 4D indices $\mu,\nu,\ldots=0,\ldots
,3$ for the observed (noncompact) dimensions; and 6D indices
$m,n\ldots=4,\ldots,9$ for the hidden (compact) dimensions.

For simplicity we restrict ourselves to geometries, $g_{mn}$,
whose volume and curvatures are all characterized by a single
scale: $M_{\KK} = 1/L$. Control over semiclassical calculations
typically requires both a small dilaton, $g_s = e^\phi \ll 1$, and
small curvatures, $\alpha'/L^2 \ll 1$, and so $M_{\KK} \ll M_s$
where $M_s^2 = 1/\alpha'$ denotes the string scale. At scales
below $M_s$ massive string modes can be integrated out, leaving an
effective-field-theory description in terms of a
higher-dimensional supergravity.

When higher-dimensional fields are dimensionally reduced on such a
space their linearized fluctuations about this background are
expanded in a complete set of Kaluza Klein (KK) eigenmodes, {\it
e.g.}
\be \label{KKReduction}
    \delta \phi(x,y) = \sum_k \varphi_k(x) u_k(y) \,,
\ee
with mode functions $u_k(y)$ chosen as eigenfunctions of various
differential operators arising from the extra-dimensional kinetic
operators, $\Delta u_k = \lambda_k u_k$. The differential operator
is chosen so that the resulting 4D fields, $\varphi_k(x)$, satisfy
the appropriate field equations for a particle having a mass
$m_k$, which is computable in terms of the corresponding
eigenvalue $\lambda_k$. This produces a tower of states having
masses ranging between 0 and $\infty$ (for marginally stable
vacua), which are split in mass by $\Delta m_k \sim O(M_{\KK})$.

The effective field theory describing energies smaller than
$M_{\KK}$ is a four dimensional one, because there is insufficient
energy to excite KK modes which can probe the extra dimensions.
Conversely, although it is possible to think of the effective
theory above $M_{\KK}$ as a  complicated 4D theory involving KK
modes and strong couplings, it is better to think of it as being
higher dimensional for several reasons, not least of which being
that it is much simpler to do so. Moreover, it is also true that
the KK modes have masses reaching right up to the UV cutoff, and
so there is never a parametrically wide gap between the UV cutoff
and the mass of the heaviest mode. Furthermore, the 4D picture
obscures higher-dimensional spacetime symmetries, like general
covariance and supersymmetry, whose presence is crucial to the
consistency of the theory.

For warped configurations the main difference is that the vacuum
metric takes the more general form
\be
    \exd s^2 = e^{2A(y)} \eta_{\mu\nu} \exd x^\mu \exd x^\nu
    +  g_{mn}(y) \exd y^m \exd y^n \,,
\ee
where the warp factor $e^{A(y)}$ varies in the extra dimensions.
The physical interpretation of the warp factor can be found by
considering the energy of a  test particle having mass $m$ and
proper velocity $u^\ssM$. If the particle is stationary at a
specific point, $y_0$, in the extra dimensions then the
normalization condition, $u^2 = -1$, implies $u^\ssM =
\delta^\ssM_0 e^{-A(y_0)}$. The four-dimensional energy of such a
particle is then $E = - m \, \xi_\ssM u^\ssM  = e^{2A(y_0)} u^0 m
= e^{A(y_0)} m$, where $\xi^\ssM = \delta^\ssM_0$ is the timelike
Killing vector field corresponding to time translation. We see
that the energy of such a test particle is highly suppressed in
strongly warped regions where $e^{A} \ll 1$.

The KK reduction of fluctuations about such a space takes a form
similar to eq.~\pref{KKReduction}, but with mode functions which
diagonalize different differential operators and satisfy different
normalization conditions than in the unwarped case. For instance,
a higher-dimensional scalar fluctuation of ten dimensional mass
$m_{10}$, satisfies the equation
\be
    (\Box_{10} - m_{10}^2)\Phi = \left[ e^{-2A}
    \eta^{\mu\nu} \partial_\mu
    \partial_\nu + \frac{e^{-4A}}{\sqrt{g}} \partial_m \left(
    \sqrt{g}e^{4A} g^{mn} \partial_n \right) - m_{10}^2 \right]
    \Phi = 0 \,.
\ee
Four-dimensional KK masses are given by eigenvalues of the operator
\be \label{KKDelta0}
    \Delta
    =  - {e^{-2A} \over \sqrt{g}} \partial_m (\sqrt{g}e^{4A}
    g^{mn}\partial_n ) + e^{2A}m_{10}^2,
\ee
rather than operator $\Delta_0 = - (1/\sqrt{g})
\partial_m (\sqrt{g} g^{mn} \partial_n )+ m_{10}^2$ which would
have been used in the absence of warping. Notice one effect of
warping is to convert the 10D mass $m_{10}$, into a potential
$e^{2A(y)} m_{10}^2$ for the wavefunctions of the KK excitations.
For a wavefunction localized in a region of large warping, with
the local warp factor $ e^{A} \ll 1 $, one can expect a redshifted
four dimensional mass $m\sim m_{10}e^{A} $ in accord with the
preceding discussion.

In type IIB string theory, it is possible to write explicit
solutions for warped compactifications \cite{GKP}. We shall use
these solutions as our laboratory to study the effects of warped
throats on supersymmetry breaking. First, we review some features
of these compactifications which shall be relevant for our
discussion.

In these constructions, the geometry takes the form
\begin{equation}
    \exd s_{10}^{2} = e^{2A} \eta_{\mu \nu} \exd x^{\mu}
    \exd x^{\nu} + e^{-2A} \tg_{mn}(y) \exd y^{m} \exd y^{n},
    \label{gkpm}
\end{equation}
where $\tg_{mn}$ is a metric on  Calabi-Yau of fiducial volume
$\tilde{V}={\alpha '}^3$.
%
With this metric the generic 4D KK masses for a scalar field
satisfying $(-\Box_{10} + m_{10}^2)\Phi = 0$ are given by the
eigenvalues of
\be \label{KKDelta}
    \Delta =  - {e^{4A} \over \sqrt{\tilde{g}}}
    \partial_m (\sqrt{\tilde{g}}
    \tilde{g}^{mn}\partial_n ) + e^{2A}m_{10}^2,
\ee
rather than eq.~\pref{KKDelta0}.

The warp factor satisfies the equation of motion
\begin{equation}
    - \tilde{\nabla}^{2} e^{-4A} =  \frac{G_{mnp}
    \bar{G}^{\widetilde{mnp}}}{12 \, \im \tau }
    + 2 \kappa_{10}^{2} T_{3}
    \tilde{\rho} _{3}^{\mathnormal{loc}}\,
    \label{wfac}
\end{equation}
where $\tau= C_0+i e^{-\phi}$ is the axio-dilaton, $G_3=F_3-\tau
H_3$ is the complex three-form field strength, the tilde indicates
indices raised with ${\tilde g}^{mn}$, and ${\tilde \rho}^{loc}_3$
represents localized sources of D3-brane charge.

Given a particular solution $e^{-4A_0}$ of (\ref{wfac}),  we can
always find a family of solutions \cite{hepth0507158} with
parameter $c$,
\be
    e^{-4A} = e^{-4A_0} + c \,.
\ee
One convenient choice is to take the particular solution to be
orthogonal to the zero mode $c$,
\begin{equation}
    \int \exd^6y \sqrt{\tilde g} \, e^{-4A_0} =0 \,,
    \label{zmorth}
\end{equation}
which emphasizes that $e^{-4A_0}$ cannot then be everywhere
positive.
This agrees with the statement that this quantity may become
negative in regions of string-size around negative tension
objects, where the supergravity approximation fails.

For large $c$, in most parts of the manifold, the warp factor is
approximately constant and the background resembles that of a
standard Calabi-Yau compactification. This Calabi-Yau like region
is often referred to as the \textit{bulk}.
In this case $c$ sets the scale of the metric over much of the
bulk and so controls the overall size of the compactification,
with
%
\begin{equation}
    V = \int \exd^6y \sqrt{g_6} \approx c^{3/2} \int
    \exd^6y \sqrt{\tilde{g}_6} = c^{3/2} \tilde{V} \,,
\end{equation}
and so $c \approx (V/\tilde V)^{2/3} \propto {\cal V}^{2/3}$,
where $V$ is the volume of the internal metric and ${\cal V}$ the
dimensionless volume relative to the string scale.

For later use, we also note the form of the metric associated with
the four-dimensional Einstein frame (which is related to
(\ref{gkpm}) by  rescaling the noncompact dimensions):
\begin{equation}
    \exd s_{10}^{2} = \lambda [e^{-4A_0} + c]^{-1/2}  \eta_{\mu \nu}
    \exd x^{\mu} \exd x^{\nu} + [e^{-4A_0} + c]^{1/2} \tg_{mn}(y)
    \exd y^{m} \exd y^{n},
\label{ef}
\end{equation}
with
%
\begin{equation}
    \frac{1}{\lambda} =  \frac{1}{\tilde{V}} \,
    \int \exd^6 y \sqrt{\tg}(c +
    e^{-4A_0}) \,, \label{lamb}
\end{equation}
chosen to ensure a canonical 4D Einstein-Hilbert action. With the
choice (\ref{zmorth}) this simplifies to
\begin{equation}
         \lambda ={1\over c} \propto {\cal{V}}^{-2/3} \ .
\end{equation}
For generality we often quote results for arbitrary $\lambda$,
bearing in mind the specialization to Einstein frame through
(\ref{lamb}).

As shown in \cite{GKP}, regions of strong warping can arise in
Type IIB vacua from typical values of the flux quantum numbers.
For instance, if $M$ units of R-R flux and $K$ units of NS-NS flux
wrap the A and B cycles of a conifold locus respectively, the
solutions to eq.~\pref{wfac} can become large, attaining a finite
maximum value
\begin{equation}
    \lb{wconifold}
       e^{-4A_m } \simeq e^{8 \pi K/3M g_{s} }\ .
\end{equation}
We define a throat as a region where the relative redshift,
$\Omega$, is particularly large compared to other points on the
manifold. Since $\Omega$ is given by the ratio of $e^{{A}}$ at the
two points in question, and because the warp factor in the bulk is
$e^A \simeq c^{-1/4}$, eq.~(\ref{wconifold}) implies a maximum
redshift
%
\begin{equation}
    \Omega_m = {(e^{-4A_{m}} + c)^{-1/4} \over c^{-1/4} }
    = \left( 1 + \frac{e^{-4A_m}}{c} \right)^{-1/4} \,,
               \label{rs}
\end{equation}
relative to the bulk. Note that this depends on the volume of the
compactification through its dependence on $c \propto {\cal
V}^{2/3}$. We have a strongly warped throat if
\begin{equation}
               c \ll e^{-4A_m} \lb{thr}
\end{equation}
and this redshift factor is large: $\Omega_m \simeq e^{A_m}c^{1/4}
\propto e^{A_m} {\cal V}^{1/6}$. The condition (\ref {thr})
moreover implies that the geometry (\ref{gkpm}) in the throat
region is largely independent of $c$ and so also of the overall
volume of the compactification. If, on the other hand,
\begin{equation}
               c \gg e^{-4A_m} \lb{lvol}
\end{equation}
then the relative redshift (\ref{rs}) tends to unity and the
geometry is that of a usual Calabi-Yau compactification.\footnote{
This is in keeping with the fact that fluxes are $\alpha'$
effects, and so {\it{all}} flux-induced effects should disappear
for sufficiently large volume. This argument also applies to
settings other than type IIB.}


Fluxes generically introduce masses for complex structure moduli
and the dilaton \cite{GKP}, regardless of whether or not they
break supersymmetry, and we pause here to describe how these scale
with ${\cal V}$ and $e^{-4A_m}$ as a warm-up for our later
discussion of the gravitino. The mass spectrum of the excitations
of the dilaton (which can be thought of as representative of modes
that acquire flux-induced masses) was studied in detail in
\cite{hepth0603233}, and is obtained by linearizing the equations
of motion \cite{hepth0507158} for the dilaton about the
backgrounds of \cite{GKP}, as described in previous sections.
Linearizing the equation of the dilaton for an excitation with
four-dimensional mass $m$ and wave-function in the internal
direction $\tau(y)$, one obtains \cite{hepth0507158} in this
way\footnote{In obtaining (\ref{tlin}) we have ignored mixing with
the fluctuations of the three form and metric. We shall use the
equation only to give a qualitative explanation of the results of
\cite{hepth0603233}, a purpose for which these fluctuations are
not important.}
\begin{equation}
                \lambda e^{4A} \tilde{\nabla}^{2} \tau (y)
                +   m^{2} \tau (y) =
                \frac{g_s}{12}\lambda e^{8A} G_{mnp}
                \bar{G}^{\widetilde{mnp}}
                \tau(y) \,. \label{tlin}
\end{equation}
Notice that the `mass' term generated by the flux,
\begin{equation}
                       \frac{g_s}{12}{ \lambda e^{8{A}}  } G_{mnp}
                      \bar{G}^{\widetilde{mnp}} \, ,    \label{myi}
\end{equation}
is not a constant but varies non-trivially over the internal
manifold.

In the bulk region, the metric $\tg_{mn}$ is of order unity, hence
$G_{mnp} \bar{G}^{\widetilde{mnp}}$ is of order $n^{2}_{f}$, where
$n_{f}$ is the flux quantum which measures the amplitude of $G$.
Using $e^{-4A} \sim c$ for the bulk, we find
\begin{equation}
            { g_s\lambda e^{8{A}}  } G_{mnp}
                      \bar{G}^{\widetilde{mnp}}    \sim \frac{
                      \lambda n^{2}_{f} }{{\cal V}^{4/3} },
             \label{bm}
\end{equation}
leading to \cite{hepth0507158,hepth0603233}
\begin{equation}
           m_{\tau} \sim \frac{n_{f}
           \sqrt{\lambda}}{{\cal {V}}^{2/3}} \,.
             \label{ntm}
\end{equation}
This should be compared with the scaling of a generic 10D or KK
mass, such as those implied by \footnote{Recall
that $\lambda \sim {\cal{V}}^{-2/3}$, $L \sim 1$ in our conventions
and in the bulk $m_{10} \sim {\cal{V}}^{-1/2}$.
 } eq.~\pref{KKDelta}:
\ba \label{GenericBulkMasses}
    M_{10} &\sim& m_{10} \sqrt{\lambda} \left(e^{-4A_m} + c
    \right)^{-1/4} \sim
    \frac{\sqrt{\lambda}}{{\cal V}^{1/6}}  \; m_{10}\nn\\
    \hbox{and} \quad
    M_{\KK} &\sim& \sqrt{\lambda} \left(e^{-4A_m} + c
    \right)^{-1/2} \, \frac{1}{L}
    \sim \frac{\sqrt{\lambda} }{ {\cal V}^{1/3}}
    \; \frac{1}{L} \,,
\ea
where $L$ is a characteristic length scale measured using the
metric $\tilde{g}_{mn}$, and the last approximate equalities in
both cases specialize to the bulk and use $c \propto {\cal
V}^{2/3}$.


In the throat region, by contrast, the underlying Calabi-Yau
metric $\tg_{mn}$ has a shrinking three cycle, and the presence of
the fluxes makes the internal manifold (with metric $e^{-2A}
\tg_{mn}$) have a characteristic length scale in this region of
the order of $ \sqrt{n'_{f}} $ in string units, where $n'_{f}$ is
an integer which quantizes the amplitude of fluxes threading the
cycles in the throat region.\footnote{  For instance in the
infrared end of the KS throat $n'_{f} \sim M$, the flux threading
the $S_{3}$ of the conifold. } Provided the wave-function of the
mode of interest is localized in this throat region, we therefore
expect $G_{mnp}\bar{G}^{mnp} \sim 1 /n'_{f}$ (and so $G_{mnp}
\bar{G}^{\widetilde{mnp}} \sim e^{-6A_m}/n'_f$, where each
factor of $\tilde{g}^{mn}$ contributes a power of
$e^{-2A_m}/n'_{f}$). Hence at the bottom of the throat we find
\begin{equation}
    { g_s\lambda e^{8{A}}  } G_{mnp}\bar{G}^{\widetilde{mnp}}
    \sim { \lambda  e^{2A_{m}} \over n'_{f} }
        \label{dtmt} \,,
\end{equation}
leading to a KK mass of order \cite{hepth0603233}
\begin{equation}
            m_{\tau} \sim e^{A_m} \sqrt{
            \frac{\lambda}{n'_{f}}}\,.
            \label{dtm}
\end{equation}
For comparison, eqs.~\pref{GenericBulkMasses} give the following
estimates for 10D masses and KK mode energies in the strongly
warped region
\ba \label{GenericThroatMasses}
    M_{10}^w &\sim& \sqrt{\lambda} \, e^{A_m}  \; m_{10}\nn\\
    \hbox{and} \quad
    M_{\KK}^w &\sim& \frac{\sqrt{\lambda} \, e^{2A_m}}{L}
    \sim \frac{\sqrt{\lambda} \, e^{A_m}} {\rho} \,,
\ea
where we again use that the cycle size measured by the metric
$\tilde{g}_{mn}$ deep inside a throat scales with the warping as
$L \sim e^{A_m} \rho$, where $\rho$ is warping independent.


One expects the dilaton to localize in the infrared end of the
throat region and acquire a mass of the order of (\ref{dtmt})
whenever it is energetically favorable to do so, {\it i.e.} when
the volume is small enough so that (\ref{dtmt}) is less than
(\ref{bm}). This yields the condition
\begin{equation}
            {\cal  V}^{2/3}
            \lsim e^{-A_m},
            \label{edis}
\end{equation}
which is equivalent to $c \lsim e^{-A_m}$. Since $c \sim {\cal
V}^{2/3} \gg 1$, eq.~\pref{thr} ensures this condition is
satisfied within a strongly warped throat.

The phenomenon of localization of massive modes should be fairly
generic. As the volume of the compactification decreases, the
redshift (\ref{rs}) becomes more and more prominent and one
expects energetics to drive the wave-function of excitations into
the throat region, ensuring the localized mode is continuously
connected to the lowest mode in the regime (\ref{lvol}). That is,
as the volume of the compactification is decreased the
wave-function of the dilaton continuously varies from being
uniformly spread throughout the internal manifold to being highly
localized in the throat. Furthermore, since the generic KK mass
gap in the strongly-warped regime (\ref{thr}) is of the same order
as the mass (\ref{dtm}), all modes of the dilaton-axion should be
integrated out of the effective field theory for the majority of
4D applications.

Our analysis of the gravitino (to follow in section 3) finds
similar effects. In the presence of SUSY-breaking flux the
wave-function of the lightest KK modes also localize into the
throat region. In what follows we shall confine our discussion to
this strongly-warped regime and examine its implications for SUSY
breaking. But first we comment on various important energy scales
and possible effective descriptions in this regime.


\subsubsection{Energy scales and effective descriptions in the
strongly warped regime} \label{ESEDSWR}


Let us be more explicit about scales in the strongly-warped
situation, with $c \lsim e^{-A_m}$. The relevant energy scales
are: the four-dimensional Planck mass $M_p$, which is the basic
scale in the Einstein frame and so convenient to set to unity.
Notice that the bulk string scale is $M_s \sim
 \sim g_s {\cal V}^{-1/2}$, and the
bulk KK scale in the Einstein frame is $M_{\KK}\sim {\cal
V}^{-2/3}$ in these units (recall eqs.~\pref{GenericBulkMasses}
with $\lambda \sim {\cal V}^{-2/3}$). Moreover, strong warping
produces the warped string scale $M_s^w \sim g_s e^{A_m}
{\cal V}^{-1/3}$ and the warped Kaluza-Klein scale, $M_{\KK}^w
\sim  g_s e^{A_m} {\cal V}^{-1/3}/\rho \sim M_s^w/\rho$ ({\it
c.f.} eqs.~\pref{GenericThroatMasses}) where $\rho
>1$ is a characteristic length of the tip of the throat (in units
of the string length). Notice that the volume dependence of
$M_{\KK}^w$ and $M_s^w$ is the same and they only differ somewhat
by the factor $\rho$. This is an important difference with respect
to the bulk quantities since this implies that the tower of string
states and Kaluza-Klein states at the tip of the throat are not
hierarchically different as ${\cal V}$ gets large.

Based on this structure of scales we can distinguish the following
energy regimes (and effective theories which describe them) in
strongly warped models.
\begin{enumerate}
\item $E < M_{\KK}^w$: This energy range is below the mass of
moduli that acquire flux induced mass and all KK and string modes.
The low-energy effective theory describing these energies is
necessarily a 4D field theory. It contains light degrees of
freedom (like the K\"ahler moduli, which are massless until
$\alpha'$ corrections and non-perturbative effects are accounted
for \cite{GKP,KKLT}).

\item $M_{\KK}^w < E <M_s^w$: This energy range (which can exist
if $\rho \gg 1$) is below the mass of all massive string states,
warped or not, but contains a (finite) tower of those KK modes
which are localized deep within the warped region. Because all
string modes are massive, they can be integrated out leaving a
low-energy effective theory which in this case is an explicitly
higher-dimensional field theory. This effective theory is {\it
not} the higher-dimensional field theory describing the full
compact space. (If it were it would include states having masses
larger than $M_{\KK}$, which is larger than the assumed cutoff.)
Instead it only probes the warped geometry, with cutoff at a point
$y = Y$ where the warp factor, $e^{A(Y)}$, is no longer
sufficiently small. This energy scale represents both an UV cutoff
(since it excludes KK and string states having energy higher than
$e^{A(Y )} / {\cal V}^{1/3}$) and an IR cutoff, since the
condition $y < Y$ gives the space a finite volume, providing an
example of UV/IR mixing.
\item $M_s^w < E < M_{\KK}$:\footnote{Since we are assuming large
warping we are taking $M_s^w<M_{\KK}$. For small warping we can
have $M_{\KK}<M_s^w$ and the regime $M_{\KK}<E<M_s^w$ is the
standard 10D supergravity, as in the usual unwarped case.} The
effective theory for this energy range contains both massive KK
{\it and} string modes, as well as non-perturbative excitations
like branes and black holes, but again only those with strongly
warped-suppressed spectra. As such, the low-energy effective
theory must be a string theory
--- again {\it not} the full string theory with which one starts,
but rather a string theory which lives only within the cut-off
volume of the warped geometry.
\item $M_{\KK} < E < M_s$: This energy range includes only
strongly warped string modes, but contains the KK modes of the
higher-dimensional field theory. The low-energy effective theory
in this case would appear to consist of the string theory
localized to the warped region (as above), but coupled to the full
set of supergravity modes which can propagate outside of the
warped region. Such a theory is somewhat novel and it would be
interesting to elucidate further its explicit description.
\item $M_s < E$: In this energy range the appropriate description
is the full string theory, defined in the entire
higher-dimensional geometry. It is believed that this theory can
apply to arbitrarily high energies.
\end{enumerate}

Cases 3 and 4 above may na\"\i vely seem unusual inasmuch as they
involve effective cut-off {\it string} theories, with the strings
propagating in nontrivial, but cut-off, higher-dimensional
background fields. They are indeed bona fide string theories since
their cutoffs are much higher than the masses of the lightest
(strongly warped) string states.  Of course, in the light of
AdS/CFT duality they seem less novel.  One expects an alternate
description of both regimes 2 and 3 above in terms of a
cutoff gauge theory with the appropriate relevant deformation
\cite{Klebanov:2000hb}. Moreover, one expects to be able to
describe regime 4 in terms of the supergravity of the bulk
manifold coupled to a cutoff gauge theory description of the
throat dynamics.  This is a variant of the description of the
effective field theory of the single brane RS scenario
\cite{RStwo} as that of a cutoff conformal field theory coupled to
four-dimensional gravity and other ultraviolet brane degrees of
freedom \cite{GiKa,ArRa}. It would be interesting, but goes beyond
the scope of this article, to further investigate properties of
such theories, and to moreover understand the space of such
possible theories and their possible deformations through
excitation of stringy modes above the threshold $M_s^w$.

We now put aside such discussion and instead ask how supersymmetry
breaking manifests itself in case 1.

\subsection{Supersymmetry breaking}

We next turn to the relevance of these various scales for
supersymmetry breaking.

\subsubsection{Supersymmetry breaking scales and warping}
\label{degenerate}

Consider, then, a higher-dimensional compactification of a
supersymmetric field (or string) theory. Although a generic
compactification will break all of the supersymmetries of the
higher-dimensional theory, we choose to focus here on
compactifications --- like those considered for Type IIB vacua in
ref.~\cite{GKP}, say --- chosen to preserve, or to approximately
preserve, one of these supersymmetries (from the 4D perspective).
`Approximate' conservation here means that the scale associated
with splittings within the various 4D supersymmetry multiplets is
sufficiently small, in a way made more precise below.

Any higher dimensional supersymmetry parameter,
$\varepsilon(x,y)$, consists of many independent supersymmetries
when viewed from the 4D perspective.  Specifically, let $a$ and
$\alpha$ be four- and six-dimensional spinor indices,
respectively, so that the combination $(a\alpha)$ serves as a
ten-dimensional spinor index.  Then $\varepsilon$ can be expanded
\be
    \varepsilon^{a\alpha}(x,y) = \sum_k \epsilon^a_k(x)
    \, \eta^\alpha_k(y) \,
\ee
in terms of an appropriate basis of 6D spinors, $\eta_k(y)$, and
each of the 4D spinors, $\epsilon_k(x)$, defines a separate local
4D supersymmetry transformation, gauged by the appropriate 4D
component of the gravitino field,
\be
    \Psi_\mu^{a\alpha}(x,y) = \sum_k \psi_{\mu k}^a(x)
    \, \eta^\alpha_k(y)
    \,,
\ee
with $\delta \psi_{\mu k} = D_\mu \epsilon_k + \ldots$ under a
supersymmetry transformation. We suppress spinor indices in the
sequel.

If precisely one 4D supersymmetry is unbroken then by assumption
one of the spinor modes, $\eta_0$, is an appropriate ``Killing
spinor," for which the corresponding gravitino mode $\psi_{\mu 0}$
is precisely massless. Above this massless state will be a KK
tower of massive 4D spin-3/2 states, $\psi_{\mu k}$, whose
lightest elements we expect to have mass $\sim M_{\KK}$ in an
unwarped environment, or $\sim M_{\KK}^w$ in a strongly warped
environment. The unbroken supersymmetry ensures that the physics
below all of these scales is captured by an effective 4D
supergravity.


Imagine now breaking this last 4D supersymmetry, perhaps by
turning on a nonzero flux, a configuration of branes or
non-perturbative effects. Then all supersymmetries are broken, and
in general there is no longer a uniquely defined gravitino which
can be identified as `the' 4D gravitino, to the extent that an
approximate 4D description is possible. Two possible equivalent
ways to define the 4D gravitino in this case are: ($i$) the
gravitino having the lightest nonzero mass, since this gauges the
4D supersymmetry which is the least broken, and ($ii$) the
gravitino which is adiabatically related to the massless gravitino
as parameters are adjusted to restore an unbroken 4D
supersymmetry.

In this section we argue that for strongly-warped systems the
spin-3/2 state to which these lead has an extra-dimensional
wave-function which is localized within the warped region. As a
consequence it generically does not couple with $M_p$-suppressed
couplings (unlike the massless 4D graviton), and so is unlikely to
be approximately described by a simple 4D supergravity. In this
case an alternative definition of the 4D gravitino is required,
and experience with supersymmetric RS models
\cite{SusyRS1,SusyRS2} suggests identifying the 4D gravitino as
the state which couples to the strongly interacting CFT in the
dual description of the strongly warped throat\footnote{We thank
  T. Gherghetta and A. Pomarol for useful discussions on these
  points.}.
 In this dual
picture all of the lightest gravitino KK modes localized in the
throat represent spin-3/2 resonances (similar to the massive KK
graviton modes), and the much-more-massive, `fundamental'
gravitino is the state whose extra-dimensional wave-function most
resembles that of the massless 4D graviton.

To study these issues in more detail, imagine turning on a
supersymmetry breaking background field that is not itself
confined to the strongly warped region, and so has an amplitude
set by a scale, $\tilde\Lambda$, that is not warped. In what
follows we take $\tilde\Lambda$ to be bounded above by $M_{\KK}$,
as would be the case for flux-generated SUSY breaking. The
quantity $f = \tilde{\Lambda} / M_{\KK} \lsim 1$ is then a small
(continuous or discrete) parameter, and one 4D supersymmetry is
restored in the limit that $f \to 0$. As a result we know that
when $f \to 0$ all particles residing within the same 4D
supermultiplet for this symmetry have precisely the same mass, and
in particular one of the 4D KK gravitino modes is precisely
massless. But for $f \ne 0$ all 4D supersymmetries are broken and
all KK mode masses are perturbed by an $f$-dependent amount which
in general splits the masses of different particles residing
within the same 4D supersymmetry multiplet.

We now ask: When $f \ne 0$ how large are the splittings, $\Delta
m_k$, within supermultiplets of the least broken 4D supersymmetry?
And in particular, how massive is the lightest 4D gravitino? In
the unwarped case, the answer for both questions would be $m_{3/2}
\sim \Delta m_k \sim f M_{\KK}$. In warped geometries masses of
the same order are generically also expected for gravitino modes
which are not localized within the strongly warped throat, such as
for the explicit truncation of the 10D gravitino to the mode which
is massless in the limit $f \to 0$ \cite{hepth0208123},
\be
    \Psi_\mu(x,y) = \psi_{\mu 0}(x) \, \eta_0(y) \,.
    \label{gravtrunc}
\ee
We emphasize that such behaviour is expected only in the neighbourhood
of the region of parameter space coresponding to $f \to 0$, away from
this limit one expects the wavefunction of the gravitino to
be modified and eventually to localize in the throat.

The next section, \ref{warp222}, shows that this same result is
{\it not} true for the least broken 4D supersymmetry in warped
geometries, whose gravitino is (by definition) the lightest
gravitino KK state. To this end we generalize our earlier
discussion for the dilaton to see how flux-induced gravitino KK
masses scale with the parameters $e^{A_m}$, $c$ and ${\cal V}$ in
warped geometries, and show that the lightest gravitino KK states
are localized in the strongly warped regions, with energies that
are characterized by the warped KK scale. In this case we argue
that the mass of the least massive gravitino is warped to smaller
values, being at most of order the warped KK scale if
$\tilde\Lambda$ is smaller than $M_{\KK}$:
\be
    m_{3/2} \sim f' M^w_{\KK} \label{mtt} \,,
\ee
provided $f' M^w_\KK \ll f M_\KK$. We quantify the size of the
supersymmetry-breaking parameters $f$ and $f'\ll 1$ below.


\subsubsection{The gravitino mass}
\label{warp222}

In this subsection we examine the gravitino equations of motion to
obtain the criterion for the localization of the gravitino
wavefunction and its mass in this regime, and to obtain estimates
for the order of magnitude of the parameters $f$ and $f'$. We
shall follow the conventions in Appendix B of \cite{hepth0208123}
for the gravitino equations of motion, with $\kappa=1$. First, we
set our gamma matrix conventions. The 10 dimensional gamma
matrices $\Gamma^{M}$ satisfy the algebra
\begin{equation}
\{ \Gamma^{M} , \Gamma^{N} \} = 2 g^{MN}\, .
\end{equation}
For the metric  (\ref{ef}), we take
\begin{equation}
 \Gamma^{{\mu}} = {e^{-\ta} \over \sqrt{\lambda} }
 \gamma^{\mu} \otimes 1  \  \ \
     \
 \Gamma^{{m}} = e^{\ta} \gamma_{c} \otimes \tilde{\gamma}^{{m}},
\end{equation}
where
\begin{equation}
               \{ \gamma^{\mu} , \gamma^{\nu} \} = 2 \eta^{\mu
               \nu} \ \ \{ \tilde{\gamma}^{m} , \tilde{\gamma}^{{n}}
               \} = 2\tilde{g}^{mn},
\end{equation}
and $ \gamma_{c} = i \gamma_{0} \gamma_{1} \gamma_{2} \gamma_{3} $
is the four dimensional chirality matrix. The gravitino equation
of motion is
\begin{equation}
\label{gravitinoeqnmotion}
 \Gamma^{MNP} \hat{D}_{N} \Psi_{P} = - { i \over 2 } \Gamma^{P}
\Gamma^{M} \hat{\lambda}^{*} P_{P} - { i 
\over 48 } \Gamma^{NPQ}
\Gamma^{M} \hat{\lambda} G^{*}_{NPQ} +{\cal O}(\Psi^3)
\end{equation}
where $\hat{\lambda}$ is the dilatino, and $P_{M}$ is the field
strength of the dilaton
\begin{equation}
  P_{M} = { 1 \over 1 - BB^{*} } \partial_{M} B\ ,
\end{equation}
with
\begin{equation}
    B = { 1 + i
    \tau \over 1 - i \tau }\ \nonumber \,.
\end{equation}
The supercovariant derivative acting on the gravitino is given by
\begin{equation}
    \hat{D}_{N} \Psi_{P} = D_{N} \Psi_{P} - 
    R_{P} \Psi_{N} -
    S_{P} \Psi_{N}^{*}
\end{equation}
with \nonumber
\begin{equation}
    R_{M} = { i \over 480 } ( \Gamma^{M_{1} ... M_{5} } F_{M_1 ....
    M_{5}} ) \Gamma_{M}
\end{equation}
and
\begin{equation}
     S_{M} =  {1 \over 96} ( \Gamma_{M}^{\ \ NPQ} G_{NPQ} - 9
    \Gamma^{NP} G_{MNP})\ .
\end{equation}

The complicated nature of the equations makes it difficult to find
an explicit solution corresponding to the massive 4D gravitino. In
fact, it is not consistent to excite just the fields $\Psi^{\mu}$;
as in the case of dilaton and complex structure
moduli \cite{hepth0507158},
one finds mixing between the various 10D supergravity modes while
carrying out the KK reduction. Given the intractable nature of the
equations of motion, we shall use energetic arguments to determine
the condition for localization of the gravitino motivated by the
observations  (\ref{myi} -- \ref{edis})  for the dilaton in
section \ref{warp211}.

We take the four dimensional gravitino $\psi_{\mu} (x)$, to be
embedded in the ten dimensional gravitino as
\begin{equation} \label{an}
    \Psi_{\mu}(x,y) = \psi_{\mu} (x) \otimes \eta (y)
\end{equation}
where  $\eta (y)$ is the wavefunction of the gravitino in the
extra dimensions. Thus,
components of the gravitino equation of motion in the non-compact
direction have the structure
\begin{equation}
     {e^{-3\ta} \over \lambda^{3/2} } \gamma^{\mu \nu \rho}
    \partial_{\nu} \psi_{\rho} \otimes \eta (y)
    + { 1 \over 24 \lambda} \gamma^{{\mu}\nu}\gamma_c
    \psi^{*}_{ \nu} \otimes e^{\ta} 
    G_{mnp} \gamma^{\widetilde{mnp}} \eta^{*} + ....  = 0
    \ ,  \label{gre}
\end{equation}
where we have not explicitly written the contributions of the
terms containing derivatives of $\eta(y)$ and other contributions
neglected due to the specific form of our ansatz. The term
\begin{equation}
    \gamma^{\mu \nu \rho} \partial_{\nu} \psi_{\rho} \lb{ko}
\end{equation}
is the standard kinetic term for a spin $3/2$ field in four
dimensions. One expects the mass to be of the order of the
relative strength of the kinetic term  and the term involving
fluxes in (\ref{gre}). We assume that the terms not written
explicitly in (\ref{gre}) scale in a similar way as a function of
the warping and Calabi-Yau volume as do the terms that are shown,
justifying their neglect in obtaining an estimate for the mass.
This should be reasonable since all such terms must adjust so that
the equations can be satisfied, and appears confirmed by a crude
analysis of their effects in the equations of motion.

Comparing the kinetic and potential terms in (\ref{gre}) we find
that the flux-induced mass term for the gravitino varies across
the internal manifold, and scales like
\begin{equation}
\label{mt}
    \sqrt{\lambda}e^{4\ta} G_{mnp} \gamma^{\widetilde{mnp}} \,.
\end{equation}
Keeping in mind that the gamma matrices scale like the ``square
root of the inverse metric,"  and comparing with (\ref{myi}),  we
see that the mass term has the same dependence on the warp factor
and the fluxes as the mass term for the dilation defined in
(\ref{myi}). Thus, we find for weak warping ($c \gsim e^{-A_m}$) a
flux-induced gravitino mass of order
\be \label{gmbulk}
    m_{3/2} \propto  \frac{ \vartheta }{{\cal V}}   \,,
\ee
where $\vartheta$ is the strength of supersymmetry breaking fluxes.
In terms of the definition $f \sim
m_{3/2}/M_{\KK}$ we therefore have
\be\label{efe}
    f \sim  \frac{ \vartheta }{{\cal V}^{1/3}} \,.
\ee

Similarly, for strong warping ($c \lsim e^{-A_m}$) the lightest
gravitino wave-function localizes, giving a mass of order
(where $\lambda \sim {{\cal{V}}^{-2/3}})$
\begin{equation}
    m_{3/2} \propto \vartheta' \, {e^{A_m}}
    \sqrt{\frac{\lambda}{n_f'}}  \,,
     \label{gm}
\end{equation}
and so $f' = m_{3/2}/M_{\KK}^w$ is
\be
    f' \sim \frac{\vartheta' \rho}{\sqrt{n_f'}} \,,
\ee
where $\vartheta'$ describes the relative strength of the SUSY
breaking and supersymmetric fluxes that thread cycles that are
localized within the throat.

To summarize, the above arguments provide two estimates for the
size of KK gravitino masses arising from supersymmetry-breaking
fluxes in warped geometries. In particular we expect those KK
modes which are not localized in the warped areas to generically
have masses which are of order $\Delta M \equiv f M_{\KK} \sim
 \vartheta /{\cal V}$ . By contrast, for strongly warped throats ({\it i.e.}
those for which ${\cal V}^{2/3} \ll e^{-A_m}$) some gravitino
modes can lower their masses to $\Delta m \equiv f' M_{\KK}^w \sim
\vartheta' e^{A_m}/\sqrt{n_f'}\, {\cal V}^{1/3}$ by localizing
within the strongly warped areas.


\subsubsection{The Supersymmetric Limit}

We include the factors $\vartheta$ and $\vartheta'$ in the above
expressions in order to capture the fact that fluxes can, but need
not, break supersymmetry. Indeed, completely supersymmetric warped
geometries with fluxes exist, and this limit is captured in the
above expressions for $m_{3/2}$ through the limit
$\vartheta \to 0$. Notice also that the parameters $f$ and $f'$
can be made systematically small, either by taking small
$\vartheta , \vartheta'$ or large ${\cal V}$ or $n_f'$. Being able
to ensure small $f$ and $f'$ is important in the next section,
where we discuss the low-energy 4D effective theory, and ask
whether the mass of the lightest gravitino is hierarchically
separated from the mass of other excitations.

Before examining the implications of these scales for the
low-energy effective 4D theory, we first pause to argue that it is
the lowest energy KK mode which is adiabatically related to the
massless gravitino in the supersymmetric limit, even though its
localization into the throat ensures that its extra-dimensional
wave-function differs from the direct supersymmetric truncation,
eq.~(\ref{gravtrunc}). At first sight this is a surprising claim,
particularly if the supersymmetry breaking parameter $f$ is small,
because the mode \pref{gravtrunc} represents the massless state
when $f \to 0$. And in perturbation theory it usually suffices to
use unperturbed eigenfunctions, $\psi_0$, in order to find the
leading-order correction to the corresponding eigenvalues, $\delta
E \sim (\psi_0, H_{\rm int} \psi_0)$, and so one might expect
estimates based on the supersymmetric gravitino wave-function to
therefore capture the leading contributions to the lightest
nonzero gravitino mass once supersymmetry breaks.

Yet we have seen that in the strongly warped regime the
wave-function of the lightest gravitino KK mode in the presence of
supersymmetry breaking fluxes is localized in the highly warped
region and so differs significantly from the Killing spinor on the
Calabi Yau. This localization is responsible for the lowering of
its mass to the warped KK scale, and is not captured by an
expression like $\delta E \sim (\psi_0, H_{\rm int} \psi_0)$ since
this does not correct the wave-function. The key point is that for
warped geometries one must use {\it degenerate} perturbation
theory, because there are many unperturbed (supersymmetric) states
having warped masses which can be much smaller than the typical
matrix element of the perturbation, $f M_{\KK}$. Because of this
degeneracy the system can minimize its energy in the presence of
the supersymmetry-breaking perturbation by choosing an appropriate
linear combination of states, which in the present instance are
the states localized within the warped throat.

If we imagine adiabatically turning on an infinitesimal
supersymmetry-breaking perturbation (or deepening the warping of a
throat), we expect non-degenerate perturbation theory to apply so
long as $f M_{\KK} \ll f' M_{\KK}^w$, and in this limit the
wave-function of the lightest gravitino KK mode should remain very
close to the supersymmetric state, eq.~\pref{gravtrunc}. However,
once the supersymmetry-breaking flux is large enough that $f
M_{\KK}$ becomes larger than $f' M_{\KK}^w$ it becomes
energetically favourable to concentrate into the throat, and so as
$\vartheta$ and $\vartheta'$ are increased the lightest gravitino
state  continuously evolves into
a localized state with a warped flux-induced KK mass. In this
sense the evolution of the gravitino wave-function with increasing
warping resembles the more familiar process of the repulsion of
atomic energy levels or the resonant oscillations of neutrinos in
matter.\footnote{We thank Henry Tye for comments on this point.}
Further discussion of the nature of the KK wave-functions in
highly warped geometries and the use of degenerate perturbation
theory in this context is given in the Appendix, where a toy
example is analyzed in more detail.


\subsection{Criteria for supersymmetric low-energy actions}

We have discussed, in section \ref{ESEDSWR}, when the low-energy
action of a system should be 4-dimensional or higher dimensional,
and how the existence of strongly warped throats complicates the
procedures which are used in unwarped situations. In this section
we similarly review the criteria for when the low-energy limit of
a higher-dimensional supergravity (or string) theory should be
described by a (possibly spontaneously broken) supersymmetric, or
explicitly non-supersymmetric effective theory.

{}From the 4D point of view, higher dimensional supergravity
theories broken by bulk flux fields are special cases of `hidden
sector' models (for a review see \cite{HSModels}). In these models
there is a collection of low-energy fields, $\ell^a$, of physical
interest (describing, say, Standard Model particles). These are
assumed to be coupled to a more generic set of fields, $h^m$,
whose dynamics somehow breaks supersymmetry. Although
supersymmetry is badly broken in the `hidden' sector described by
$h^m$ (with typical supermultiplet mass splittings of order
$\Delta M$), the weak $h-\ell$ couplings ensure this is only
weakly transmitted to the `light' sector described by $\ell^a$
(whose supersymmetry breaking splittings are $\Delta m \ll \Delta
M$). In the best-case scenario these couplings are gravitational
in strength. Notice that there is no requirement that the
hidden-sector fields be light, although there is typically one
state in this sector, the goldstone fermion,\footnote{Or, more
precisely, the massive gravitino which `eats' it.} which is much
lighter than the others.

The form of the low-energy effective field theory which describes
this kind of situation below a UV cutoff, $\Lambda$, depends in an
important way on the relative size of $\Lambda$, $\Delta M$ and
$\Delta m$, as follows.
\begin{enumerate}
\item $\Delta M \ll \Lambda$: If the cutoff is larger than all
supersymmetry-breaking mass splittings then the field content of
the low-energy theory can be grouped into supermultiplets. In this
case the low-energy theory is itself described by a supergravity,
even though it may include some of the hidden-sector fields. Any
spontaneous supersymmetry breaking within the full theory can be
understood within the effective theory as spontaneous
supersymmetry breaking due to the appearance of a SUSY-breaking
v.e.v. purely within the effective theory.
\item $\Lambda \ll \Delta m$: If the cutoff is well below the
smallest SUSY-breaking scale, then a generic supermultiplet has
some elements which are heavier than $\Lambda$ and so are
integrated out, while others are lighter than $\Lambda$ and so
remain in the low-energy theory. In this case the field content of
the low energy theory cannot be organized into supermultiplets (it
might contain just fermions with no bosons, for example), and so
supersymmetry must be {\it nonlinearly realized} \cite{NLSUSY}.
(Notice that for a gauge theory a nonlinearly-realized spontaneous
breaking is operationally indistinguishable from explicit breaking
within the low-energy theory below the breaking scale \cite{BL},
and so in this case the effective theory can be an arbitrary
non-supersymmetric field theory.)
\item $\Delta m \ll \Lambda \ll \Delta M$: If the cutoff lies
between the two splitting scales, then there generically are
supermultiplets in the hidden sector (split by $\Delta M$) for
which some particles are heavier than $\Lambda$ and so are
integrated out, while others are lighter than $\Lambda$ and so
remain in the low-energy theory. This is true in particular for
the multiplet which contains the goldstone fermion in the hidden
sector. In this case supersymmetry is generically badly broken in
the effective theory. What distinguishes this case from case 2
above, is that supersymmetry breaking is much smaller than
$\Lambda$ within the light sector, which therefore has the field
content to fill out complete supermultiplets. As a result, {\it
provided we restrict our attention only to light-sector
observables} the breaking of supersymmetry in this sector can be
described as a supergravity coupled to a collection of
soft-breaking terms\footnote{By `soft breaking' we do not here
mean terms which do not generate quadratic divergences, but
instead have in mind the more general usage, spelled out in more
detail below, in which supersymmetry breaking arises through terms
obtained by replacing hidden-sector auxiliary fields by their
SUSY-breaking expectation values.} \cite{SoftBreaking,BI} which
encode the couplings to supersymmetry breaking in the hidden
sector.
\end{enumerate}

For extra-dimensional supergravity without warping, the low-energy
theory of interest is often defined to cover energies smaller than
the KK scale, $M_{\KK}$, and so consists of the effective 4D
interactions of the various KK zero modes. In this case we can
regard the hidden sector to consist of the massive KK modes and
the light sector to consist of the low-energy KK zero modes. If
supersymmetry is broken by extra-dimensional physics, such as by
fluxes, then the low-energy 4D theory below the KK scale in
general need not be supersymmetric. If, however, the mass
splitting, $\Delta m \sim f M_\KK$, among the KK zero modes
satisfies $\Delta m \ll M_{\KK}$ (and so $f \ll 1$), the above
discussion shows that the low-energy theory can be an effective 4D
supergravity (possibly coupled to soft-breaking terms which
capture the effects of integrating out parts of badly-split KK
supermultiplets).



In the case with strong warping, we can take the light sector to
consist of those modes which are localized within the strongly
warped region, whose masses are typically of order $M_{\KK}^w$ or
smaller. The hidden sector consists of those fields which are not
so localized. We have seen that supersymmetry-breaking splittings
in the light sector are of order $\Delta m = f' M_{\KK}^w $ while
those in the hidden sector can be much larger, of order $\Delta M
= f M_{\KK}$. (Figure \ref{scales} sketches some of the relevant
scales for the light modes in the two cases, $\Delta m \ll \Delta
M \ll M^w_\KK$ and $\Delta m \ll M^w_\KK \ll \Delta M$.) A
complicating feature in this case is that a 4D description is not
valid at all unless a hierarchy exists between the states of
interest and the generic warped KK scale $\Lambda \ll M_{\KK}^w$,
and this generically need not be the case if all of the light
gravitino states are split by $\Delta m \sim f' M^w_\KK$, which is
of the same order as the lightest gravitino mass.

\FIGURE{\centering%
\includegraphics[width=3.8in]{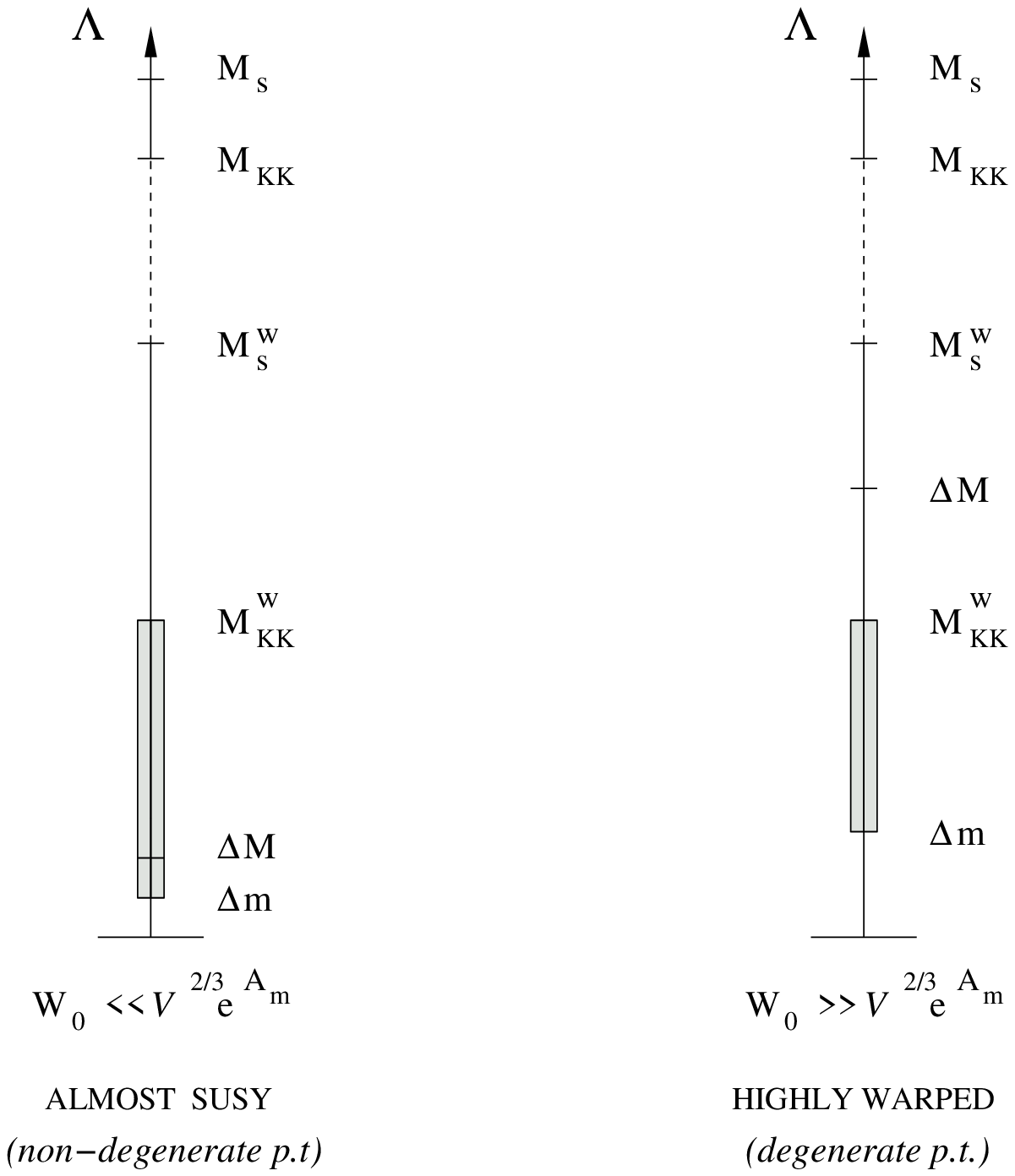}
\caption{Relevant scales for the localized modes at the infrared
bottom of a throat. In the case $W_0 \ll {\cal{V}}^{2/3}
e^{A_{min}}$, the effects of the fluxes are almost negligible and
the system is nearly supersymmetric. For ${\cal{V}}^{2/3}
e^{A_{min}}\gg W_0$, the effects of the fluxes however become
important. The masses of the unperturbed states $M^w_{\KK}$ are
much smaller than the flux scale and non-degenerate perturbation
theory no longer applies to this case. The shaded region shows the
range of energies at which an ultraviolet cutoff gives rise to a
4D supergravity description for the light modes.} \label{scales}}

As a result, in the strongly warped case the absence of a clean
hierarchy between the scale of supersymmetry breaking and the
onset of the extra-dimensional description indicates we should
generically not expect to obtain a low-energy description
consisting purely of a 4D supergravity, even supplemented by soft
supersymmetry breaking terms. Conversely, if a single gravitino
were much lighter than the others, its interactions would be well
described by a 4D supergravity, and in particular would
necessarily be suppressed by the 4D Planck scale, $M_p$, as are
those of its superpartner, the massless 4D graviton. However, the
interactions of localized gravitino states in strongly warped
regions are generically suppressed at low energies only by a
warped scale, such as $M^w_\KK$, and so are generically much
stronger than are those of the massless 4D graviton.

These considerations suggest instead formulating the effective 4D
theory in terms of the dual variables, in terms of which the
semiclassical geometrical degrees of freedom in the throat are
described by some sort of strongly interacting conformal
field theory (CFT). In this picture the stronger interactions of
the tower of warped KK gravitons and gravitini are regarded as
expressing the relatively strong residual interactions amongst
resonant spin-3/2 and spin-2 bound states of the CFT constituents,
with the corresponding supersymmetries being emergent symmetries
associated with the strong CFT interactions.

The presence of a strongly-coupled CFT makes the criteria for
describing this dual system in terms of a 4D supergravity more
complicated to formulate. However a generic situation for which
further progress is possible is in the case where we choose not to
directly measure properties of the CFT and instead integrate it
out and ask for its implications on other low-energy states (like
the 4D graviton, bulk moduli, and so on). Provided the couplings
between these low-energy modes and the CFT are sufficiently weak
the CFT can simply act as a hidden sector, whose implications
might be captured by a (possibly softly broken) effective 4D
theory. Because the 4D gravitino which would be relevant to such a
theory would have couplings similar to the 4D graviton, it could
be expected to be a bulk state whose mass is of order $f M_\KK$,
and so considerably heavier than the warped KK scales.
This more complicated situation is in the spirit taken by much of
the model building within supersymmetric 5D RS models
\cite{SusyRS1,SusyRS2}.

\subsubsection{Effective 4D supersymmetry}
\label{SSUnbroken}

Suppose first we consider case 1, for which all supersymmetry
breaking scales are small compared with the warped KK scale. (We
might imagine this is arranged, for instance, by appropriately
choosing the parameter $\vartheta$.) In this case we expect the
low energy limit to be given by a 4D effective supergravity,
described, as usual, by a K\"ahler potential, $K(\ell,\ell^*)$, a
superpotential, $W(\ell)$, and a gauge kinetic function,
$F_{AB}(\ell)$. For instance, leading order calculations in Type
IIB flux compactifications give
\be\label{lk}
    K = -2 \ln {\cal V} \,, \qquad \hbox{and} \qquad
    W = W_0 \,,
\ee
where ${\cal V}$ is the volume of the underlying Calabi-Yau space
expressed as a function of its holomorphic moduli, and $W_0$ is
the Gukov-Vafa-Witten superpotential \cite{GVW}, regarded as a
function of the holomorphic complex structure moduli.

If all supersymmetries are broken by the fluxes in the underlying
compactification, $N=1$ 4D SUSY will be spontaneously broken in
the 4D effective theory, leading to a mass for the 4D gravitino of
the form
\be
    m_{3/2} = e^{K/2} \, |W| = \frac{|W_0|}{{\cal V}} \,,
\ee
in the absence of strong warping. This uses the lowest-order
results $K = -2 \ln {\cal V}$ and $W = W_0$.  We see that the freedom
to  dial fluxes to small
values in the microscopic theory to obtain small supersymmetry
breaking effects is described in the low-energy theory by the
freedom to choose very small values for $W_0$. Also, comparing this
with (\ref{gmbulk}) we find the that in the regime that there
is a four dimensional effective description, $\vartheta \sim |W_{0}|$.


As discussed above, in the case of strong warping it is not
generic that the low-energy theory is well described by a 4D
supergravity at all. However, it can happen that a 4D supergravity
can apply, even if there exist strongly warped regions. One such a
case arises in the supersymmetric limit, for which there is always
a clear hierarchy between the massless gravitino and the warped or
unwarped KK scales. Suppose the low-energy matter sector in this
case consists of a strongly warped throat. The theory then
contains warped states, whose low-energy interactions with the
supergravity sector can be of interest.

In such a case we would expect an effective 4D supergravity which
must know about the warped scale, and we now ask how this might be
encoded into the low-energy theory. Since the influence of fluxes
in general arise at higher order in $\alpha'$, the modifications
to the low-energy supergravity which encode this warping might be
expected to arise as $\alpha'$ corrections, and so in particular
arise as modifications to $K$. (In principle the explicit form for
this warping dependence could be captured by matching to the
underlying Calabi-Yau dynamics, such as was done for 5D
supersymmetric RS models in ref.~\cite{KWarp}, but calculations
are hampered in the 10D case by not knowing the explicit form for
the underlying geometry in this case.)

Part of the answer of how warping can enter $K$ to describe states
that are localized in the throat is given by adding $2A_m$ to $K$,
since this provides an appropriate overall scaling down of all
masses (with fixed 4D Planck mass),
\be
    K(\ell, \ell^*) = 2A_m + { \check K}(\ell, \ell^*) \,,
    \label{kscale}
\ee
with the holomorphic functions $W$ and $F_{AB}$ unchanged. Such a
constant piece does not contribute at all to the K\"ahler metric,
$\partial_a \partial_{\overline a} K$, or to the covariant
derivative, $D_a W = \partial_a W + W \partial_a K$, but does have
the effect of scaling the scalar potential by an overall factor:
$U = e^{2A_m} \check U$, where $\check U$ is the scalar potential
computed using the K\"ahler potential $\check K$. In particular,
the entire scalar mass matrix gets scaled by a corresponding
amount, $m^2 = e^{2A_m} \check{m}^2$. Fermion masses also get
scaled down in an identical way, due to the ubiquitous factor of
$e^{G/2}$ to which they are proportional, with
\be
    G = K + \ln W + \ln W^*\, .
    \label{gdef}
\ee
The 4D gravitino does not share this warping in the supersymmetric
limit because it is massless, by assumption. Such a constant
contribution to $K$ also appears as a leading term in a
large-volume expansion of the K\"ahler function identified for
supersymmetric RS models \cite{KWarp}.

\vspace{1.5cm}

\subsubsection{4D action with softly broken supersymmetry}
\label{SSSoftlyBroken}

We next turn to the case of more direct phenomenological interest,
case 3, wherein we obtain our effective theory by integrating out
a supersymmetry-breaking sector but keep the goldstone fermion
whose mixing with the 4D gravitino gives it its mass. In this case
we expect that if the SUSY-breaking sector couples sufficiently
weakly to the observable sector, then its supersymmetry-breaking
effects can be encoded in the low-energy theory by a suitable
class of weakly-coupled soft-breaking interactions.

A warped example which is described by this type of soft breaking
is obtained by perturbing our earlier supersymmetric picture of a
supersymmetric throat. Suppose we now turn on a
supersymmetry-breaking flux in the throat, but not by enough to
localize the lightest gravitino KK mode. More precisely, suppose
$f' M^w_\KK \gsim f M_\KK$ even though $M^w_\KK \ll M_\KK$, as
would be possible if $\vartheta /\vartheta' \lsim e^{A_m} {\cal
V}^{2/3} \ll 1$ (and so, in particular, if $ \vartheta \to 0$). In
this situation it does not energetically pay the lightest
gravitino to localize, allowing it to have $M_p$-suppressed
couplings\footnote{We could also entertain the possibility of $f'\ll
  f$ and a light gravitino localized in the warped region. While this
  is an interesting scenario, an explicit calculation supporting
  the existence of such a regime  is not available at present.}.
 Yet its mass is warped because this sets the scale of
the SUSY-breaking physics in the warped throat.

In this case we expect the low-energy theory below the bulk KK
scale to be a 4D supergravity coupled to a strongly interacting
supersymmetric CFT describing the throat. Alternatively,
integrating out the CFT can lead to a softly-broken 4D
supergravity describing the remaining, observable, sector provided
only that the couplings to the CFT are sufficiently weak. We now
show that the generic warping of the resulting
supersymmetry-breaking masses may also be accomplished by the same
shift as for the supersymmetric case, $K = 2A_m + \check K$.

To show this we require a statement of what the resulting
soft-breaking interactions might be. These have been enumerated in
the literature \cite{SoftBreaking,BI}, under the assumption that
there is a regime for which the full theory --- both $\ell^a$'s
and $h^m$'s --- is given by a 4D $N=1$ supergravity, described by
an appropriate K\"ahler potential,\footnote{Although in the
previous section the hidden fields $h$ included moduli and matter
fields living far from the throat, here they are understood to
include only moduli which may survive at low energies (such as
K\"ahler moduli).}
\be
    K = \hat{K}(h,h^*) + \tilde{K}_{a\overline{b}}(h,h^*) \ell^a
    \ell^{\overline{b}} + \frac12 \, Z_{ab}(h,h^*) \ell^a \ell^b
    + \cdots \,,
\ee
as well as a superpotential, $W$, and gauge-kinetic function,
$F_{AB}$. (Here $\ell^{\overline{a}}$ denotes $(\ell^a)^*$.) In
this case the soft-breaking quantities can be computed in terms of
the assumed coupling functions and the SUSY-breaking hidden-sector
auxiliary fields,
\be
    \cF^m = e^{G/2} K^{m\overline{n}} \partial_{\bar{n}} {G}
    \,, \label{auxF}
\ee
using eq.~(\ref{gdef}). For instance, the resulting expressions
for the scalar and gaugino masses are \cite{BI}
\bea \label{SoftBreakingMasses}
    m^2_{a\overline{b}} &=& (m^2_{3/2} + V_0)
    \tilde{K}_{a\overline{b}}
    - \cF^m \cF^{\overline{n}} \Bigl( \partial_m
    \partial_{\overline{n}}
    \tilde{K}_{a\overline{b}} - \tilde{K}^{c \overline{d}}
    \partial_m
    \tilde{K}_{a\overline{d}} \partial_{\overline{n}}
    \tilde{K}_{c
    \overline{b}} \Bigr) \nn\\
    M_{AB} &=& \frac12 \, \left[\left(\hbox{Re} \,
    F \right)^{-1}\right]_{AC}
    \cF^m \partial_m F_{CB} \,,
\eea
where as before the gravitino mass is
$m_{3/2} 
= e^{K/2} |W|$
%
and $V_0$ is the value of the potential at its minimum.

We again expect the low-energy theory to know that all of the
nonzero masses associated with localized states within the warped
region are suppressed by a common factor $e^{A_m}$. As may be seen
from the above formulae, this can be done if the K\"ahler
potential contains an additive constant, which can be taken to be
in the part describing the hidden sector
\be
    \hat{K}(h,h^*) = 2A_m + {\cal K}(h, h^*) \,,
\ee
for essentially the same reason as for the supersymmetric case
considered above. Such a constant has the effect of scaling all of
the supersymmetry breaking v.e.v.'s, $\cF^m$, by a common factor
of $e^{A_m}$, thereby ensuring that all masses are properly
suppressed by the warp factor.


Similarly, using eqs.~\pref{efe} in the expression for the mass of
the 4D gravitino itself gives
\be
    m_{3/2} \sim e^{A_m} \, e^{{\check K}/2} \, W_0
    \sim \frac{\vartheta \, e^{A_m}}{{\cal V}^{1/3}} \,,
\ee
This can be used as a guide to determine the
K\"ahler potential in the regime of strong warping.

\section{SUSY breaking in the microscopic theory}
\label{warpsec3}

In this section we  give examples of how visible-sector
supersymmetry-breaking  masses might arise within a microscopic
calculation within a warped environment. To this end we compute
the warp-factor dependence of the masses which are induced for
some brane moduli as a consequence of bulk fluxes.

\subsection{Flux-induced masses on a D3-brane}
\label{warpsec31}

We consider a D3-brane filling the non-compact dimensions of a
generic GKP vacuum \cite{GKP}. The matter content of the theory in
the world-volume of the brane will contain six real scalars $Y^m$
parameterizing the position of the D3-brane in the transverse
space. The dynamics of these scalars is described in terms of the
corresponding Dirac-Born-Infeld (DBI) and Chern-Simons (CS)
actions
\be
  S_3 = -|\mu_3| \int d^4x \; e^{-\phi}\sqrt{-\det(P[E])}
  + \mu_3\int P[C_4]+\dots \, ,\label{s3}
\ee
where $\mu_3$ is the D3-brane charge, and $P[E]$ denotes the
pullback to the brane of the tensor $E_{MN}=g_{MN}+B_{MN}$. With
brane coordinates $(x^\mu, Y^m(x^\mu))$, this can be written
\be
    P[E]_{\mu\nu} = E_{\mu\nu} + E_{mn} \partial_\mu
    Y^m \partial_\nu Y^n + \partial_\mu Y^m E_{m\nu}
    + \partial_\nu Y^n E_{\mu n}
    \,. \label{pullback}
\ee
$P[C_4]$ similarly denotes the pull-back of the RR 4-form
potential. One expects additional terms in the CS piece due to the
other RR fields, but these turn out to be irrelevant for the
purpose of computing the scalar masses, as one may check. The
metric in these expressions is the string-frame metric, which is
related to (\ref{ef}) by an $e^{\phi/2}$ scaling factor,
\be
    \exd s^2_{str} = e^{\phi/2} \left( \lambda e^{2A(y)}
    \eta_{\mu\nu} \, \exd x^\mu \exd
    x^\nu + e^{-2A(y)}  \tilde{g}_{mn} \, \exd y^m
    \exd y^n\right) \,,
\ee
with $\lambda$ defined  in (\ref{lamb}).

Assuming a constant background for the ten dimensional dilaton, as
generically happens in the absence of D7 branes, one easily shows
that
\be
    \sqrt{-\det(P[E])}=\lambda^2e^{4A}e^{\phi}\left[1+
    \frac{1}{2}\lambda^{-1}e^{-4A}\tilde{g}_{mn}\partial_{\mu}
    Y^m\partial^{\mu}Y^n\right]
\label{dete}
\ee
where we keep only the terms relevant for the computation of the
scalar masses and where the internal metric $\tilde{g}_{mn}$ in
(\ref{dete}) is evaluated at the position of the brane. From this
we find
\be
    S_3\simeq -\int d^4 x \left[ |\mu_3|{\lambda\over 2}
    {\tilde g}_{mn} \partial_\mu Y^m\partial^\mu Y^n +
    |\mu_3|\lambda^2 e^{4A} -\mu_3 C_{0123}\right]\ .
    \label{d3act}
\ee
Thus the potential for a D3 position $Y^m$ is given by
\be
    V = \lambda e^{4A} - {1\over\lambda} C_{0123}\ .
\ee
This can be expanded around the minimum value $e^{4A_m}$ to give
the mass matrix,
\be
    V\simeq \lambda e^{4A_m} + {\partial_m\partial_n V\over 2}
    Y^m Y^n\ .
\ee
The trace of the mass matrix can be computed (see e.g.
\cite{hepth0209200,hepth0311241}) using the supergravity equations
\cite{GKP}:
\be
    {\tilde g}^{mn} \partial_m\partial_n V =
    \lambda {\tilde \nabla}^2 \left( e^{4A} -
    {1\over \lambda^2} C_{0123}\right)
    \simeq \lambda{e^{8A}\over 24 \, \im \tau}
    (G_3 + i{\tilde *}_6 G_3)_{mnp} ({\bar G}_3
    - i {\tilde *}_6 {\bar
    G}_3)^{\widetilde{mnp}}\ . \label{trmass}
\ee

The fact that the scalar masses vanish for imaginary self dual
fluxes can be traced back to the no-scale structure of these Type
IIB vacua~\cite{hepth0208123,dellis}. However, imaginary anti-self
dual components for the 3-form flux can be thought of as being
induced by the backreaction of effects which break the no-scale
structure. Comparing (\ref{trmass}) to (\ref{dtmt}), we find that
the flux-induced scalar masses evaluated in a warped throat are of
the same size as both the bulk moduli masses discussed there and
the gravitino mass (\ref{gm}),
\be
    m_Y\sim e^{A_m} \sqrt{\lambda} \ . \label{d3sup}
\ee

Notice that, with very little effort, the computation can be
extended to the case of the scalar masses for a $\overline{D3}$
brane, as its action differs only in the sign of the last term in
(\ref{d3act}). Thus, the expression (\ref{d3sup}) remains valid
for an antibrane, but now the left hand side of (\ref{trmass}) is
proportional to the imaginary self dual components of the
fluxes~\cite{hepth0311241}.

\subsection{Flux-induced masses on a D7 brane}

For D7 branes one can proceed as in the previous section. However,
now the D7 brane wraps a 4-cycle $\Sigma$ in the extra dimensions.
We consider a trivial normal bundle for the embedding of $\Sigma$
in the compact manifold, so there are just two geometric moduli
$Y^i$ in the four dimensional theory parameterizing the position
of the D7-brane. We use 4D indices $\alpha,\beta,\ldots$ for the
compact dimensions within the brane volume; and 2D indices
$i,j,\ldots$ for the compact dimensions transverse to the
D7-brane. If we imagine the D7 to wrap directions 4,5,6,7 of the
compact space so that 8,9 denote the transverse directions, then
$\alpha,\beta,\ldots =4,\ldots ,7$ and $i,j,\ldots =8,9$.

For simplicity, we only consider the reduction of the DBI piece of
the action, which for a D7-brane reads,
\be
    S_7 = -|\mu_7| \int_{\mathbb{R}^4\times\Sigma} \exd^8
    \xi \; e^{-\phi} \sqrt{-\det(P[E])} \,.
\ee
Notice that, as for D3-branes, one expects also contributions from
the CS action, cancelling part of the mass terms in the DBI
piece.\footnote{See \cite{hepth0408036} for further details.}
However, for the purpose of computing the warp suppression of the
scalar masses, it suffices to analyze the DBI contributions.

The pull-back (\ref{pullback}) is then given by
\bea
    P[E]_{\mu\nu} &=& e^{2A} \lambda e^{\phi/2} \eta_{\mu\nu} +
    e^{-2 A} e^{\phi/2}
    \tg_{ij} \partial_\mu Y^i \partial_\nu Y^j \nn\\
    P[E]_{\alpha\beta} &=& e^{-2A} e^{\phi/2}
    \tg_{\alpha\beta} +B_{\alpha\beta} \,,
\eea
where we restrict ourselves to the case $B_{\mu\nu} = 0$ and
$\partial_\alpha Y^i = 0$. The two fields $Y^i(x)$ denote
low-energy fluctuations in the transverse position, $y^i =
Y^i(x)$, of the D7 brane.

Because of its block-diagonal structure, the determinant of $P[E]$
becomes
\bea
    \det(P[E]) &=& (e^{2A} \lambda e^{\phi/2})^4
    \det\Bigl[ \eta_{\mu\nu}
    + e^{-4A} \lambda^{-1} \tg_{ij} \partial_\mu
    Y^i \partial_\nu Y^j
    \Bigr] \nn\\
    &&\qquad \times (e^{-2A} e^{\phi/2})^4 \det
    \Bigl[\tg_{\alpha\beta}
   + e^{2A} e^{-\phi/2} B_{\alpha\beta} \Bigr] \,,
\eea
which, when expanded in powers of the fluctuations gives
\bea
    \sqrt{-\det(P[E])} &=& \lambda^2 e^{2\phi} \sqrt{{\tilde g}_4}
    \left( 1 + {1\over2} e^{-4A} \lambda^{-1} \tg_{ij}
    \partial_\mu Y^i
    \partial^\mu Y^j + \cdots\right) \nn\\
    && \qquad\qquad\qquad \times \left( 1 + {1\over4} e^{4A}
    e^{-\phi} B_{\alpha\beta} B^{\widetilde{\alpha\beta}}
    + \cdots\right)
    \,,
\eea
where we denote  $\det \tg_{\alpha\beta}$, the determinant of the
pullback on the four cycle, by $ \tg_{4} $. We can use this in the
DBI action, working to quadratic order in the fluctuations; we
assume that the flux $B_{\alpha \beta}$ is constant over the four
cycle, and hence it does not induce Freed-Witten anomalies
\cite{hepth9907189} in the worldvolume of the brane. This gives
\bea
    S_7 &=& -|\mu_7| \int_{\mathbb{R}^4\times\Sigma}
    \exd^4 x \, \exd^4 y \, \sqrt{{\tilde g}_4}
    \; e^{\phi}
    \left(\lambda^2 + {1\over2} e^{-4A} \lambda
    \tg_{ij} \partial_\mu Y^i
    \partial^\mu Y^j + {1\over4} e^{4A} \lambda^2 e^{-\phi}
    B_{\alpha\beta} B^{\widetilde{\alpha\beta}}
    + \cdots\right) \nn\\
    &=& - |\mu_7| \int \exd^4 x  \, \left( V
    + {1\over 2}  G_{ij} \partial_\mu Y^i
    \partial^\mu Y^j\right)
\eea
where the potential and kinetic terms are given by
\be
    V=\lambda^2 \int_\Sigma d^4y  \sqrt{{\tilde g}_4}
    e^\phi\left[1+{1\over 4} e^{-\phi} e^{4A} B_{\alpha\beta}
    B^{\widetilde{\alpha\beta}} \right]\ ,\ G_{ij} = \lambda
    \int_\Sigma d^4 y \sqrt{{\tilde g}_4}e^{-4A} e^\phi {\tilde
    g}_{ij}\ .
\ee

We are now in a position to see why the presence of a nonzero
background flux, $H_{mnp} \ne 0$, can generate a mass for the
brane modulus $Y^i$. Notice that since $\exd B = H_3$, if the
normal to $\Sigma$ has components along the cycle supporting
$H_3$, in the vicinity of $\Sigma$ we may write $B_{\alpha\beta}
\sim H_{\alpha\beta i} Y^i$.  This gives
\be
    B_{\alpha\beta} B^{\widetilde{\alpha\beta} }
    \sim H_{\alpha\beta i}
    {H^{\widetilde{\alpha\beta}}}_j Y^i Y^j \,,
\ee
leading to a potential matrix $ V_{ij} \sim C \, \lambda^2
H_{\alpha\beta i} {H^{\widetilde{\alpha\beta}}}_j$. A mass is
produced if the D7 brane wraps a cycle which overlaps the cycle
which supports the nonzero flux.

Whether such masses break 4D supersymmetry or not depends on
the details of the fluxes involved. Using a complex basis in
the compact dimensions, this mass can preserve supersymmetry if
the corresponding flux is purely of (2,1) or (1,2) type, and
breaks supersymmetry \cite{GP} if it contains fluxes of (3,0) or
(0,3) type.

Since our interest is in tracking how masses depend on warping for
states localized in warped regions, it is instructive to
specialize the above analysis to the special case where the D7
brane is localized within such a region,\footnote{For an example
of such a construction, see \cite{QuevUrang}.} analogously to our
discussion of D3 branes. Suppose then that the warp factor is
$e^{A(y) }= e^{A_m} \, e^{\cA(y)}$ throughout $\Sigma$, where
$e^{A_m} \ll 1$ and $e^{{\cA(y)}}$ integrates over the cycle to
give a result which is $O(1)$ (notice that this implies that the
flux relevant for generating masses for D7 moduli is also nonzero
in the highly warped region, which is not the generic case). In
this case, using the explicit formula for the warped throat metric
we find fluctuation $G_{ij} \sim \lambda e^{2A_m}$ and $V_{ij}
\sim \lambda^2 e^{4A_m}$. Thus the mass for the fields $Y^i$ is
\be
    m_Y \sim e^{A_m} \sqrt{ \lambda } \sim
    \frac{e^{A_{m}}}{{\cal{V}}^{1/3}}
  \, ,
\ee
which coincides with the ones obtained in the previous section for
the geometric moduli of a D3 brane.

\subsection{Effective 4D description}

We now record what would be the corresponding description of such
flux-induced masses for the brane geometric moduli from the point
of view of the low-energy 4D theory. As emphasized earlier, this
is possible if the SUSY breaking fluxes are chosen to be
sufficiently small that it does not pay the lightest KK gravitino
to become localized into the throat. Because we explicitly
consider fluxes localized in a warped throat, the supersymmetry
breaking sector lives in the warped region and the SUSY breaking
scale is warped. We consider in turn the cases where the flux
preserves and breaks supersymmetry, and these correspond to the
two cases discussed above in sections \ref{SSUnbroken} and
\ref{SSSoftlyBroken}. Our main interest is in how the overall warp
suppression factor, $e^{A_m}$, appears in the low-energy theory.

\subsubsection{Unbroken supersymmetry}

If the relevant flux does not break $N=1$ 4D supersymmetry then
the effective 4D theory is described by the standard supergravity
lagrangian, within which the D-brane moduli are represented by
complex scalars residing within chiral supermultiplets. More
precisely, the D7 moduli are described by a single complex scalar
$Y$, whereas the six geometric D3 moduli are arranged into three
complex scalars, $\tilde{Y}^a$, with $a=1,2,3$. We will discuss
both kind of moduli under the same context.

The appropriate choice of K\"ahler potential is found by
inspecting the kinetic terms for the D3 and D7 moduli, leading to
the form $K = 2A_m + \cK$, with
\be \label{CurlyK}
    \cK =  K_c(\phi,\phi^*) + K_Y(\phi,\phi^*) Y^* Y +
    K_{\tilde{Y}^a}(\phi,\phi^*)\tilde{Y}^{a*}\tilde{Y}^a +
    \cdots \,,
\ee
where $\phi$ collectively denotes all the other moduli and the
ellipses denote terms involving higher powers of $\tilde{Y}^a$,
$Y$ and their complex conjugates. We include an overall constant
$2A_m$, as was argued above to be required in order to generically
warp all nonzero masses for localized states by a factor
$e^{A_m}$. This additive factor does not affect the $\tilde{Y}^a$
and $Y$ kinetic terms, however, so agreement with the microscopic
calculation requires we also choose
\be
    K_{Y} = |\mu_7| \, \, k(\phi,\phi^*) \, \quad \quad
    K_{\tilde{Y}^a} = |\mu_3| \, \, k^a(\phi,\phi^*)\,,
\ee
with no $A_m$ dependence. For the present purposes the functions
$K_c$, $k$ and $k^a$ could be arbitrary, although they are known
explicitly for specific types of compactifications
\cite{hepph9812397,hepth0409098,hepth0501139}.

In this case the superpotential is {\it not} given only by the GVW
form, since it also acquires a dependence on the D7 moduli
\cite{hepth0408036,hepth0501139} (see also
\cite{hepth0406092,hepth0410074}), which we expand to lowest order
in $Y$:
\be
    W = W_{GVW}(\phi) + \frac{\mu_7}{2} \,
    w (\phi) \, Y^2 + \cdots \,.
\ee
As for D7 branes in the unwarped regions, we will assume that $w$
does not carry any factors of the small quantity $e^{A_m}.$

The corresponding K\"ahler derivatives become
\bea
    &&D_{\tilde{Y}^a} W = \partial_{\tilde{Y}^a} W
    + W \partial_{\tilde{Y}^a} K =
    \mu_3 k^a \tilde{Y}^{a*} W_{GVW} + \cdots \,,\\
    &&D_Y W = \partial_Y W + W \partial_Y K =
    \mu_7 \Bigl[ w \, Y
    + k \, Y^* W_{GVW} \Bigr] + \cdots \,,
\eea
which show that $\tilde{Y}^a=Y= 0$ does not break supersymmetry
(or perturb the vacuum away from vanishing potential $V$).

In this case, keeping in mind the no-scale nature of the
low-energy theory which ensures that $W_{GVW} = V=0$ at the
minimum, the scalar mass term for the brane moduli is given by the
contribution
\be
    V = e^{K} \sum_{\Phi=Y,\tilde{Y}^a}\frac{
    |D_\Phi W|^2}{K_\Phi} + \cdots
    = |\mu_7| \, e^{2A_m} e^{\cK} \frac{\left|w \, Y
    \right|^2}{k} + \cdots\,,
\ee
Notice that the $Y$ mass term now scales in the same way as it
did in the microscopic computation, whereas on the other hand, the
D3 moduli remain massless, in agreement with the no-scale
structure of the potential.

\subsubsection{Broken supersymmetry}
Next consider the case where the mass-generating flux breaks
supersymmetry. Since the supersymmetry-breaking field is not
within the low-energy theory, and since the mass splitting
generated is much smaller than the generic KK mass, in this
instance we expect the effective 4D theory to be described by a 4D
supergravity supplemented by soft-breaking terms.

In this case the $Y$ and $\tilde{Y}^a$ kinetic terms are unchanged
from the supersymmetric case, and the additive term for $K$ is
also required to ensure that all generic masses are warped by a
factor of $e^{A_m}$, and so $K = 2A_m + \cK$, with $\cK$ given by
eq.~\pref{CurlyK}. The additional suppression of the D7 modulus
mass is then described by the relevant soft-breaking mass terms in
the scalar potential. Specializing the result given in
eq.~\pref{SoftBreakingMasses} to the case of a diagonal metric,
and canonically normalizing the kinetic terms, gives the following
result for the physical $Y$ mass \cite{OldSoftSUSY}
\be
    m^2_Y =  m_{3/2}^2
     - \cF^m \cF^{\overline n} \partial_m \partial_{\overline n}
     \ln K_Y \,, \label{mysq}
\ee
where we use $V_0 = 0$, and as before $m_{3/2} = e^{K/2} |W|$ and
$\cF^m$ is given by (\ref{auxF}). Notice that both terms of
(\ref{mysq}) include a factor of $e^{2A_m}$, so that the
soft-breaking mass of $Y$ indeed has a factor of $e^{A_m}$.

Regarding the D3 moduli $\tilde{Y}^a$, due to the no-scale
structure of the scalar potential for these vacua, the gravitino
mass is generically related to $K_{\tilde{Y}^a}$ in such a way
that
\be
m_{3/2}^2
     = \cF^m \cF^{\overline n} \partial_m \partial_{\overline n}
     \ln K_{\tilde{Y}^a}\label{noscaled3}
\ee
and the soft masses for the D3 brane moduli vanish, even though
supersymmetry is being broken.

In general the no-scale structure of the scalar potential is
however spoiled by both $\alpha'$ corrections and non-perturbative
corrections to the superpotential, and the relation
(\ref{noscaled3}) will not hold. In that case, the soft masses
(\ref{SoftBreakingMasses}) for the $\tilde{Y}^a$ scalars become
\be
    m^2_{\tilde{Y}^a} =  V_0+m_{3/2}^2
     - \cF^m \cF^{\overline n} \partial_m \partial_{\overline n}
     \ln K_{\tilde{Y}^a} \,,
\ee
where we have taken again a diagonal metric and canonically
normalized kinetic terms. The value of the potential, $V_0$, now
is also generically different from zero and given by the formula
\be
V_0 = e^K\left( |DW|^2 -3|W|^2\right)
\, .
\ee
Thus, from the scaling $e^K\propto e^{2A_m}$ we find that in
scenarios with broken no-scale structure the soft masses for the
D3 moduli are also suppressed by a $e^{A_m}$ factor, in agreement
with the results obtained in section \ref{warpsec31} from a
microscopic point of view.

\section{Towards Phenomenology}
\label{warpsec4}


Clearly our results may have interesting phenomenological
implications. The study of soft supersymmetry breaking in the KKLT
scenario \cite{nilles,largev} has not included the effects of
warping, while existing studies of warping effects in
phenomenology break SUSY through brane boundary conditions rather
than fluxes (see e.g. \cite{GhPom}). Warping has only been
considered in the mechanism for lifting to de Sitter space, but in
this scenario we expect the standard model to appear from D-branes
wrapping non-trivial cycles of the Calabi-Yau manifold, and a
natural possibility is that these D-branes lie in a
strongly-warped region (for explicit constructions in this
direction see \cite{botomup,QuevUrang}). This was actually part of
the original motivation in \cite{GKP}, since the redshift in a
throat is a natural mechanism to solve the hierarchy problem if
the standard model sector lives on the tip of the throat. If the
TeV scale supersymmetry breaking throat is regarded in the dual
picture as a strongly-interacting CFT, then this can be regarded
as a stringy realization of Witten's proposal for understanding
the gauge hierarchy using dynamical supersymmetry breaking
\cite{EWDSSB}.

The substantial red-shifting due to the warp factor dependence of
$M_s^w$ implies that we can consider different scenarios depending
on the corresponding warped string scale which can take any value
between the electroweak scale ($1$ TeV) and the GUT scale ($\sim
10^{17}$ GeV). On the other hand, all the other relevant scales
are also suppressed by the same warp factor (and bulk volume
dependence). As mentioned before, we expect the warped
Kaluza-Klein scale $M_{\KK}^w$ to take values somewhat smaller
than $M_s^w$ due to the characteristic curvature scale $1/\rho$ of
the tip of the throat. Furthermore, the gravitino mass and all
soft supersymmetry breaking terms are further reduced by factors
controlling the amplitude of the supersymmetry-breaking flux,
$\vartheta, \vartheta'$. Choosing these quantities very small,
such as by localizing the sources of supersymmetry breaking into
the warped throat allows the gravitino mass to be hierarchically
smaller than $M_{\KK}^w$. Generically the lightest gravitino is
then localized in the throat, and describes a resonance in the
dual CFT and is not naturally well-described by a low-energy 4D
supergravity. However, if the couplings of the SUSY breaking
physics in the throat to the gravitino are sufficiently weak, it
may not pay the lowest mass gravitino KK mode to localize, and a
low-energy supergravity action with standard soft supersymmetry
breaking terms can be obtained. Notice that when a 4D
supersymmetric description is possible, small fluxes imply a small
value for the effective superpotential $W_0$. (Although a very
small $W_0$ is also required in the original KKLT scenario, this
is for a very different reason: to have tree-level and
non-perturbative contributions to the potential to compete, and to
justify neglecting of the perturbative corrections to the K\"ahler
potential.)

For $M^w_\KK \sim M^w_s$ at the GUT scale, we require $f' \sim
10^{-13}$ to get a TeV gravitino. (An unwarped proposal which
resembles this scenario is discussed in \cite{nilles}.) It is not
clear that such scales can be understood in a controlled
approximation, however, because the conditions $e^{-4A_m}\gg {\cal
V}^{2/3}\gg 1$ and $M_s^w/M_p \sim e^{A_m}/{\cal V}^{1/3} \sim
10^{2}$ then require a relatively small volume. Smaller warped
string scales arise more naturally in our scenario, including two
potentially attractive possibilities. Having the warped string
scale at the intermediate scale, $M_s^w\sim 10^{11}$ GeV, could
permit warped realizations of intermediate-scale string scenarios
\cite{IntScale}, which are also attractive from the point of view
of some string inflationary models \cite{CosmoApps}. Getting a TeV
gravitino mass in such models typically requires $f' \sim
10^{-7}$, which puts $W_0$ in the range of validity of the KKLT
approximations.

Alternatively, if the warped string scale were of order $M_s^w\sim
10$ TeV then $f' \sim 1/10$, would still justify the use of
effective field theory. Since statistically speaking a very small
value of $W_0$ is not preferred, one might argue that having such
a low warped string scale is a more natural scenario to have. This
would indicate a very interesting phenomenological scenario with a
very small string scale and approximately supersymmetric effective
action with soft SUSY breaking terms:
\be
    M_{1/2}\sim m_0\sim A
    \sim m_{3/2}\sim \frac{1}{10} M_s^w \sim 1 \, {\rm TeV}
\ee
As mentioned earlier, this can be considered as a stringy
realization of the dual of the dynamical supersymmetry breaking
\cite{EWDSSB} approach to the hierarchy problem, with the
geometrical picture of the exponential hierarchy being obtained
from the warped geometry, rather than from strongly coupled
dynamics in the dual conformal field theory.

For $W_0 \sim 1$ we know that the KKLT approximations fail.  In a
large class of models \cite{bbcq} an exponentially large volume
stabilization is obtained by including perturbative corrections to
the K\"ahler potential. But for very large volumes the effects of
warping are less and less important because the condition
$e^{-4A_m}\gg {\cal V}^{2/3}$ becomes more difficult to satisfy.
Interestingly enough, in our set-up, values of $W_0 \gsim 1$ would
imply a collapse of the supersymmetric field theory approximation
since the gravitino mass would be as heavy as the KK and string
modes. In this case we do not expect an effective supersymmetric
4D action to play any role and we may have to consider directly an
effective string theory phenomenology with distinctive signatures
as compared with standard effective field theories, such as the
presence of towers of KK and string states, etc.  Alternatively,
one might describe this situation in terms of the AdS/CFT dual.

One might foresee further scenarios developing depending on the
location and the source of supersymmetry breaking as compared to
the standard model.

\section{Conclusions}
\label{warpsec5}

In this paper we give a first step towards understanding the
effective description of broken supersymmetric theories in
strongly warped throats. Warped compactifications are very natural
in IIB string theory \cite{GKP,DD,MH} and provide a very rich,
local and stringy scenario to discuss supersymmetry breaking with
potentially different properties from standard scenarios of
supersymmetry breaking in terms of gravity, gauge and anomaly
mediation.

A typical compactification may have many throats and depending on
the structure of the fluxes on each throat, the physics in each of
the throats can be very different. Therefore within one single
compactification we may have local models that feel differently
the scale of supersymmetry breaking and therefore different
structure of soft-breaking terms. In a large class of models this
may not be describable in terms of effective
4D supergravities since the gravitino mass will be degenerate with
the string and KK scales.

If the flux superpotential is small enough, a natural hierarchy is
generated between the KK scale and the gravitino mass, justifying
a supersymmetric effective field theory treatment. The
exponentially large warp factor is a natural source of hierarchy
as in the Randall-Sundrum model. In our case it provides the
exponentially small scale of SUSY breaking instead of dynamical
SUSY breaking.


Our investigation allowed us to solve a puzzle regarding what
defines `the' gravitino of the low-energy 4D description. The
natural notion from the extra-dimensional point of view is the
lightest KK mode of the higher-dimensional gravitino, since this
gauges the least broken 4D supersymmetry in the problem. This is
also the state which is adiabatically linked to the massless
gravitino in the supersymmetric limit. However, in strongly warped
geometries this state prefers to be localized deep within the
warped throat, and as a result generically couples to other states
with interactions that are suppressed by the warped scale, rather
than the 4D Planck scale. Such a state can be understood as a
spin-3/2 resonance of the strongly interacting CFT which provides
the dual description of the throat. Indeed it is generically not
possible to capture the low-energy limit of such a system purely
with a 4D (possibly softly broken) supergravity.

However there are situations where this is possible, most notably
when supersymmetry breaking is localized within the throat but
couples weakly enough that it does not make it energetically
worthwhile for the lightest gravitino to localize there. In this
case the 4D gravitino is most easily identified in the dual theory
by perturbing away from the supersymmetric limit. Our analysis
also leads us to suggest that the effective K\"ahler potential for
such a system is shifted by the warp function $A_m$ at the tip of
the throat, but a general interpolation between the regimes of
weak and strong warping is unknown.

Finally, we identify several potentially interesting
phenomenological scenarios depending on the amount of warping and
the tuning of the flux superpotential $W_0$. Further investigation
of the detailed phenomenology of these new scenarios is certainly
desirable. We also expect potential applications for cosmology,
and in particular for inflationary scenarios depending on the
existence of warping.

\section*{Acknowledgements}

We thank L. Ib\'a\~nez for collaboration in the early stages of
this work. We acknowledge useful conversations on the subject of
this paper with S. Abdussalam, J. Conlon, D. Cremades, O. DeWolfe,
S. Hartnoll, S. Kachru, P. Kumar, S. Thomas, E. Silverstein,  A.
Sinha, and H. Verlinde. One of us (AM) would like to especially
thank A. Frey for many useful discussions. CB's research is
supported in part by funds from Natural Sciences and Engineering
Research Council of Canada, the Killam Foundation and McMaster
University. SBG and AM acknowledge the support of the Department
of Energy under Contract DE-FG02-91ER40618. The work of PGC was
supported by the EU under the contracts MEXT-CT-2003-509661,
MRTN-CT-2004-005104 and MRTN-CT-2004-503369. KS wishes to thank
Trinity College, Cambridge for financial support. C.B. thanks the
kind hospitality of the Galileo Galilei Institute, where some of
this work was done. SdA, SBG, FQ thank KITP Santa Barbara and the
organizers of the workshop on `String Phenomenology' for the same
reasons, and acknowledge the partial support of the National
Science Foundation under Grant No. PHY99-07949. FQ is partially
funded by PPARC and a Royal Society Wolfson award. SdA wishes to
thank the Perimeter Institute and DOE grant No.
DE-FG02-91-ER-40672 for partial support.

\appendix
\section{Warping and degenerate perturbation theory}

In this appendix we use the example of a bulk scalar in the
Randall-Sundrum scenario \cite{Goldberger} as a toy model to
discuss some features of KK reduction in highly warped regions.
The KK reduction can be carried out explicitly in this model, thus
the model is helpful for building intuition. We use it to examine
the common practice of identifying supersymmetry-breaking mass
shifts by truncating the KK reduction using supersymmetric
configurations, and show when this can be justified in terms of a
perturbative analysis. We show in particular why this analysis
breaks down in the presence of strong warping.

\subsection*{Explicit diagonalization}

Reference \cite{Goldberger} considered a massive 5-d scalar
\begin{equation}
   S = \int \ d^{5} x  \sqrt{g} \big( -\partial_{M} Y \partial^{M} Y - M^{2}
   Y^{2} \big) \label{action}
\end{equation}
in a finite domain of AdS (of radius $R$)
\begin{equation}
    ds^{2} = \frac{r^{2}}{R^{2} } \eta_{\mu \nu}dx^{\mu}
    dx^{\nu} + \frac{R^{2}}{r^{2}} dr^{2}, \ \ \ r_{0} < r < R \label{rsmetric}
\end{equation}
with Neumann boundary conditions at both ends.

Before discussing the KK reduction, we note that the above model
has the essential features to capture the dynamics of the
gravitino in IIB constructions in the highly warped regime ($ c
\ll e^{-A_{m}}$). In the throat regions of IIB constructions, the
metric typically factorizes to an $AdS_{5} \times X_{5} $
structure, for some manifold $X_5$, and so the wavefunction has a
product structure. Thus, to understand the effects of warping it
suffices to truncate to a 5D model. Also the ten dimensional mass
term for the gravitino, $ e^{3A} G_{mnp} \gamma^{\widetilde{mnp}}
$ is approximately a constant in the throat region\footnote{For a
discussion see section \ref{warp222} and related discussion of the
dilaton mass term in section \ref{warp211}.} and typically scales
as the inverse of the local AdS radius. Thus the simplest model to
consider is that of a minimal massive scalar (\ref{action}) with
$MR \sim 1$.

We give a brief outline of the results of explicit Kaluza-Klein
reduction (for details see \cite{Goldberger}). For dimensional
reduction, consider the 5D equations of motion for the ansatz
\begin{equation}
             Y (x,r) = \sum_n u_{n}(r) \phi_{n} (x)
\end{equation}
with $  \partial_{\mu} \partial^{\mu} \phi_{n} = m_{n}^{2}
\phi_{n}$ ($m_{n}$ are the four dimensional masses) . This yields
an equation for $u_{n}$
\begin{equation}
  -{1\over R^{4} r} \partial_{r}  (r^{5} \partial_{r} u_{n}) +
{ r^{2} \over R^{2} }M^{2}u_{n} = m_{n}^{2} u_{n}  \ .\label{slp}
\end{equation}
A general solution to the differential equation (\ref{slp}) can be
written in terms of Bessel functions of order $\nu = \sqrt { 4 +
(MR)^{2} }$,
\begin{equation}
       u_{n}(r) =  { N_{n} \over r^{2} } \left[ J_{\nu}\left( {
       m_{n}R^{2} \over r} \right) + b_{n} Y_{\nu}\left({
       m_{n}R^{2} \over r}\right) \right] \label{wf}
\end{equation}
where $N_{n}$ and $b_{n}$ are constants. The masses $m_{n}$
and the constants $b_{n}$ are fixed by the boundary conditions at
$r_{0}, R$. For large warping (${ r_{0} /R } \ll 1 $) one finds
that the masses are determined by the equation
\begin{equation}
 2J_{\nu} (x_{n})+ x_{n} J'_{\nu}(x_{n}) = 0
 \label{spectral}
\end{equation}
where  $ x_{n} = { R^2 \over r_{o}} m_{n}$.
In the regime $ RM \sim
1 $, the lowest root of (\ref{spectral}) is of the order of unity
and
\begin{equation}
      m_{0} \sim { r_{0} \over R } M  \label{mzero}.
\end{equation}
The wavefunction (\ref{wf}) is highly localized in the region
close to $r_{0}$.

\subsection*{Perturbative analysis}

It is illustrative to examine this result in the context of
perturbation theory, since it provides a toy example which shares
many of the features which commonly arise when supersymmetry is
broken in an extra-dimensional model. In supersymmetric models it
is often useful to imagine turning on supersymmetry breaking in a
parametrically small way, such as by turning on a small flux. In
this case our interest is often in the size of the SUSY-breaking
mass splittings which arise, as computed perturbatively in the
SUSY-breaking parameter. This kind of problem has an analogue in
the present example in the limit of small $M$, since the model
acquires a new symmetry in this limit corresponding to shifts of
the form $Y \to Y + \epsilon$. We therefore now consider computing
the scalar mass in the small-$M$ limit, in order to compare the
results obtained perturbatively with the exact results found
above. The perturbative methods that we discuss can also have
applications to situations where an explicit diagonalization is
not possible.

To proceed we treat the term involving the five dimensional mass
$M$ in (\ref{slp}) as a perturbation. We take the masses and
wavefunctions of the unperturbed ``hamiltonian" to be $\mu_{n}$
and $v_{n}$, i.e
\begin{equation}
-{1\over R^{4} r} \partial_{r}  (r^{5} \partial_{r} v_{n})  \equiv
H_{0} v_{n} = \mu_{n}^{2} v_{n} \label{up}.
\end{equation}
In the absence of the five dimensional mass term, the lowest mode
is massless $(\mu_{0} = 0)$ and has a constant wavefunction. For
the higher modes, the masses and wavefunctions $( \mu_{i}, v_{i} \
i> 0 )$ are given by (\ref{wf}) and (\ref{spectral}) with $ \nu =
2 $. We note that the first roots of (\ref{spectral}) are of  the
order of unity, hence the scale of the unperturbed KK tower is
\begin{equation}
    m_{\KK} \sim  { r_{0} \over R^{2} }.  \label{ascale}
\end{equation}

To set up the perturbative computation, we begin by introducing an
inner product
\begin{equation}
             \langle a|b\rangle  \ = \int^{R}_{r_{0}} dr \ r a(r) b(r)
\end{equation}
under which the unperturbed Hamiltonian, $H_{0}$, for the KK
states is hermitian. We normalize $v_{n}$ as
\begin{equation}
             \langle v_{n}|v_{m}\rangle \ = \delta_{nm}.  \label{norm}
\end{equation}
Given the perturbation Hamiltonian
\begin{equation}
   H' \equiv { r^{2} \over R^{2} }M^{2},
\end{equation}
the mass of the lowest mode in first order non-degenerate
perturbation theory is
\begin{equation}
          m_{0}^2 = \langle v_{0}| H' | v_{0} \rangle.  \label{ndpt}
\end{equation}
With the normalization (\ref{norm}),
\begin{equation}
          v_{0} \sim  { 1 \over R}.
\end{equation}
Then (\ref{ndpt})  gives
\begin{equation}
     m_{0} \sim M  \label{incor}.
\end{equation}
This is the analogue of supergravity calculation which estimates
the lowest gravitino KK mass by evaluating the
supersymmetry-breaking action using the Killing spinor which
defines the wavefunction of the mode which is massless in the
supersymmetric limit.

Note that the perturbative result, eq.~(\ref{incor}), is much
larger than the correct value of the mass of the lowest mode
(\ref{mzero}). First order non-degenerate perturbation theory
fails because the strength of the perturbation is large compared
to  the KK scale mass (\ref{ascale}). Under such a circumstance
one expects the wavefunction of the lowest excitation ($u_{0}$) to
be significantly different from the lowest mode in the absence of
the perturbation due to mixings between the zero mode ($v_{0}$)
and the KK modes ($v_{i}$) introduced by the perturbation. These
mixings are not captured by first order non-degenerate
perturbation theory which does not incorporate the corrections to
the wavefunction.

In situations where the KK scale is small compared to the strength
of the perturbation, a better perturbative tool is {\it{degenerate
perturbation}} theory, since effectively the KK modes ($v_{i}$)
and the zero mode ($v_{0}$) are degenerate compared to the scale
of the perturbation. Typically, one has to include a large number
of KK modes to obtain quantitatively reliable results. Since our
purpose here is to be illustrative we will include just one mode.
We shall see that this is sufficient to reproduce  correct
estimates.

For the purposes of estimate, we approximate the first KK mode by
its power law behavior; then with the normalization condition
(\ref{norm})
\begin{equation}
        v_{1} \sim { r_{0}^3 \over r^{4} }.
\end{equation}
In the subspace spanned by $v_{0}$  and  $v_{1} $, the
perturbation $H'$ has approximate matrix elements
\begin{equation}
\label{abcde}
 \left( \begin{array}{cc}
M^2  &   \ \ \  { r_{0}^3 \over R^3 }\ln(R/r_0)  M^2 \\
\\
 { r_{0}^3 \over R^3 }\ln(R/r_0) M^2  & \ \ \ { r^{2}_{0} \over R^{2} } M^2\\
\end{array}
\right).
\end{equation}
The lowest eigenvalue of the perturbation matrix
(\ref{abcde}) is of the order of ${ r^{2}_{0} \over R^{2} }
M^{2}$, which is in agreement with (\ref{mzero}).


\begin{thebibliography}{99}

\bibitem{sethi1}
K.~Dasgupta, G.~Rajesh and S.~Sethi, {\it M theory, orientifolds and
G-flux,} JHEP {\bf 9908} (1999) 023, [hep-th/9908088].

\bibitem{GKP}
S.~B.~Giddings, S.~Kachru and J.~Polchinski, {\it Hierarchies from
fluxes in string compactifications,} Phys. Rev. {\bf D66}, 106006
(2002), [hep-th/0105097].



\bibitem{PPApps}
%
  S.~Dimopoulos, S.~Kachru, N.~Kaloper, A.~E.~Lawrence and E.~Silverstein,
  {\it Generating small numbers by tunneling in multi-throat  compactifications,}
  Int.\ J.\ Mod.\ Phys.\ A {\bf 19} (2004) 2657,
  [hep-th/0106128];
%
G. Cacciapaglia, C. Csaki, C. Grojean, M. Reece and J. Terning,
{\it Top and Bottom: a Brane of Their Own,} [hep-ph/0505001].

\bibitem{CosmoApps}
S.~Kachru, R.~Kallosh, A.~Linde, J.~M.~Maldacena, L.~McAllister
and S.~P.~Trivedi,
  {\it Towards inflation in string theory,}
  JCAP {\bf 0310}, 013 (2003),
  [hep-th/0308055];
%
  X.~Chen,
  {\it Multi-throat brane inflation,}
  Phys.\ Rev.\ D {\bf 71}, 063506 (2005),
  [hep-th/0408084];
%
  X.~Chen,
  {\it Inflation from warped space,}
  JHEP {\bf 0508}, 045 (2005),
  [hep-th/0501184];
%
  M.~Alishahiha, E.~Silverstein and D.~Tong,
  {\it DBI in the sky,}
  Phys.\ Rev.\ D {\bf 70}, 123505 (2004),
  [hep-th/0404084];
  %
  C.~P.~Burgess, J.~M.~Cline, H.~Stoica and F.~Quevedo,
  {\it Inflation in realistic D-brane models,}
  JHEP {\bf 0409} (2004) 033,
  [hep-th/0403119];
%
  O.~DeWolfe, S.~Kachru and H.~L.~Verlinde,
  {\it The giant inflaton,}
  JHEP {\bf 0405}, 017 (2004),
  [hep-th/0403123];
  %
  J.~J.~Blanco-Pillado {\it et al.},
  {\it Racetrack inflation,}
  JHEP {\bf 0411} (2004) 063,
  [hep-th/0406230];
  %
  {\it Inflating in a better racetrack,}
  [hep-th/0603129];
%
  D.~Cremades, F.~Quevedo and A.~Sinha,
  {\it Warped tachyonic inflation in type IIB flux compactifications and the
  open-string completeness conjecture,}
  JHEP {\bf 0510}, 106 (2005),
  [hep-th/0505252];
%
  S.~E.~Shandera and S.~H.~Tye,
  ``Observing brane inflation,''
  JCAP {\bf 0605} (2006) 007
  [arXiv:hep-th/0601099];
%
   G.~Shiu and B.~Underwood,
  ``Observing the geometry of warped compactification via cosmic
inflation,''
  [arXiv:hep-th/0610151].

\bibitem{KKLT}
  S.~Kachru, R.~Kallosh, A.~Linde and S.~P.~Trivedi,
  {\it De Sitter vacua in string theory,}
  Phys.\ Rev.\ D {\bf 68}, 046005 (2003),
  [hep-th/0301240].

\bibitem{OldSoftSUSY}
A.~Brignole, L.~E.~Ib\'a\~nez and C.~Munoz,
{\it Soft supersymmetry-breaking terms from supergravity and superstring models,}
[hep-ph/9707209].

\bibitem{RSone}
L.~Randall and R.~Sundrum, {\it A large mass hierarchy from a small
extra dimension,} Phys.\ Rev. Lett. {\bf 83} ({1999}) {3370},
[hep-ph/9905221].
%
\bibitem{RStwo}
L.~Randall and R.~Sundrum, {\it An alternative to compactification,}
Phys. Rev. Lett. {\bf 83} ({1999}) {4690}, [hep-th/9906064].

\bibitem{HVer}
  H.~L.~Verlinde,
  {\it Holography and compactification,}
  Nucl.\ Phys.\ B {\bf 580}, 264 (2000)
  [hep-th/9906182].

\bibitem{KlSt}
  I.~R.~Klebanov and M.~J.~Strassler,
   ``Supergravity and a confining gauge theory: Duality cascades and
  {\it chiSB-resolution of naked singularities,}
  JHEP {\bf 0008}, 052 (2000)
  [hep-th/0007191].


\bibitem{BBC}
  N.~Barnaby, C.~P.~Burgess and J.~M.~Cline,
  {\it Warped reheating in brane-antibrane inflation,}
  JCAP {\bf 0504} (2005) 007,
  [hep-th/0412040].

\bibitem{WarpedReheat2}
    L. Kofman and P. Yi, {\it Reheating the Universe after
    String Theory Inflation,} [hep-th/0507257];
%
    D. Chialva, G. Shiu and B. Underwood,
    {\it Warped Reheating in Multi-Throat Brane Inflation,}
    [hep-th/0508229];
%
    X. Chen, S.-H. Tye, {\it Heating in Brane Inflation and Hidden Dark
    Matter,} [hep-th/0602136];
%
    P. Langfelder, {\it On Tunnelling In Two-Throat Warped
    Reheating,} [hep-th/0602296].


\bibitem{kim}
K.~W.~Choi, D.~Y.~Kim, I.~W.~Kim and T.~Kobayashi,
  {\it Supersymmetry breaking in warped geometry,}
  Eur.\ Phys.\ J.\ C {\bf 35} (2004) 267,
  [hep-ph/0305024];
F.~Brummer, A.~Hebecker and M.~Trapletti,
  {\it SUSY breaking mediation by throat fields,}
  Nucl.\ Phys.\ B {\bf 755} (2006) 186,
  [hep-th/0605232].


\bibitem{Goldberger}
  W.~D.~Goldberger and M.~B.~Wise,
  {\it Bulk fields in the Randall-Sundrum compactification scenario,}
  Phys.\ Rev.\ D {\bf 60}, 107505 (1999),
  [hep-ph/9907218].


\bibitem{hepth0507158}
  S.~B.~Giddings and A.~Maharana,
  {\it Dynamics of warped compactifications and the
  shape of the warped
  landscape,}
  Phys.\ Rev.\ D {\bf 73}, 126003 (2006),
  [hep-th/0507158].


\bibitem{hepth0603233}
A.~R.~Frey and A.~Maharana, {\it Warped spectroscopy: Localization
of frozen bulk modes,} [hep-th/0603233].



\bibitem{SusyRS1}
  R.~Altendorfer, J.~Bagger and D.~Nemeschansky,
  {\it Supersymmetric Randall-Sundrum scenario,}
  Phys.\ Rev.\  D {\bf 63} (2001) 125025
  [hep-th/0003117].

\bibitem{SusyRS2}
  A.~Falkowski, Z.~Lalak and S.~Pokorski,
  {\it Supersymmetrizing branes with bulk in five-dimensional
  supergravity,}
  Phys.\ Lett.\  B {\bf 491} (2000) 172
  [hep-th/0004093];
%
  {\it Five-dimensional gauged supergravities with universal hypermultiplet  and
  warped brane worlds,}
  Phys.\ Lett.\  B {\bf 509} (2001) 337
  [hep-th/0009167].


\bibitem{GhPom}
  T.~Gherghetta and A.~Pomarol,
  {\it Bulk fields and supersymmetry in a slice of AdS},
  Nucl.\ Phys.\  B {\bf 586} (2000) 141
  [hep-ph/0003129];
%
  {\it A warped supersymmetric standard model},
  Nucl.\ Phys.\ B {\bf 602}, 3 (2001)
  [arXiv:hep-ph/0012378];
%
  {\it A Stueckelberg formalism for the gravitino
  from warped extra  dimensions},
  Phys.\ Lett.\  B {\bf 536} (2002) 277
  [hep-th/0203120];
%
  {\it The standard model partly supersymmetric},
  Phys.\ Rev.\ D {\bf 67}, 085018 (2003)
  [arXiv:hep-ph/0302001].


\bibitem{KWarp}
  M.~A.~Luty and R.~Sundrum,
  {\it Hierarchy stabilization in warped supersymmetry,}
  Phys.\ Rev.\  D {\bf 64} (2001) 065012
  [hep-th/0012158];
%
  J.~Bagger, D.~Nemeschansky and R.~J.~Zhang,
  {\it Supersymmetric radion in the Randall-Sundrum scenario,}
  JHEP {\bf 0108} (2001) 057
  [hep-th/0012163];
%
  A.~Falkowski, Z.~Lalak and S.~Pokorski,
  {\it Four dimensional supergravities from five dimensional brane
  worlds,}
  Nucl.\ Phys.\  B {\bf 613} (2001) 189
  [hep-th/0102145].


\bibitem{hepth0208123}
O.~DeWolfe and S.~B.~Giddings,
{\it Scales and hierarchies in warped compactifications and brane worlds,}
Phys.\ Rev.\ D {\bf 67}, 066008 (2003),
[hep-th/0208123].

\bibitem{deAlwis:2003sn}
  S.~P.~de Alwis,
  {\it On potentials from fluxes,}
  Phys.\ Rev.\ D {\bf 68}, 126001 (2003),
  [hep-th/0307084].
  %
    S.~P.~de Alwis,
  {\it Brane worlds in 5D and warped compactifications in IIB,}
  Phys.\ Lett.\ B {\bf 603}, 230 (2004),
  [hep-th/0407126].

\bibitem{KKA}
K.~Koyama, K.~Koyama and F.~Arroja, {\it On the 4D effective
theory in warped compactifications with fluxes and branes},
  Phys.\ Lett.\ B {\bf 641}, 81 (2006)
  [hep-th/0607145].

\bibitem{Klebanov:2000hb}
  I.~R.~Klebanov and M.~J.~Strassler,
  {\it Supergravity and a confining gauge theory: Duality cascades and
  chiSB-resolution of naked singularities,}
  JHEP {\bf 0008}, 052 (2000),
  [hep-th/0007191].

\bibitem{GiKa}
  S.~B.~Giddings and E.~Katz,
  {\it Effective theories and black hole production in warped
  compactifications,}
  J.\ Math.\ Phys.\  {\bf 42}, 3082 (2001),
  [hep-th/0009176].

\bibitem{ArRa}
  N.~Arkani-Hamed, M.~Porrati and L.~Randall,
  {\it Holography and phenomenology,}
  JHEP {\bf 0108}, 017 (2001),
  [hep-th/0012148].

\bibitem{HSModels}
  H.~P.~Nilles,
  {\it Supersymmetry, Supergravity And Particle Physics,}
  Phys.\ Rept.\  {\bf 110}, 1 (1984).

\bibitem{NLSUSY}
  J.~Bagger and J.~Wess,
  {\it Partial Breaking Of Extended Supersymmetry,}
  Phys.\ Lett.\ B {\bf 138}, 105 (1984);
%
J.~Bagger and A.~Galperin,
  {\it Matter couplings in partially broken extended supersymmetry,}
  Phys.\ Lett.\ B {\bf 336}, 25 (1994),
  [hep-th/9406217];
%
  I.~Antoniadis, H.~Partouche and T.~R.~Taylor,
  {\it Spontaneous Breaking of N=2 Global Supersymmetry,}
  Phys.\ Lett.\ B {\bf 372}, 83 (1996),
  [hep-th/9512006];
%
 S.~Ferrara, L.~Girardello and M.~Porrati,
  {\it Spontaneous Breaking of N=2 to N=1 in Rigid and Local Supersymmetric
   Theories,}
  Phys.\ Lett.\ B {\bf 376}, 275 (1996),
  [hep-th/9512180];
%
 C.~P.~Burgess, E.~Filotas, M.~Klein and F.~Quevedo,
  {\it Low-energy brane-world effective actions and partial supersymmetry
   breaking,}
  JHEP {\bf 0310}, 041 (2003),
  [hep-th/0209190].

\bibitem{BL}
  C.~P.~Burgess and D.~London,
  {\it On anomalous gauge boson couplings and loop calculations,}
  Phys.\ Rev.\ Lett.\  {\bf 69} (1992) 3428;
  %
  {\it Uses and abuses of effective Lagrangians,}
  Phys.\ Rev.\ D {\bf 48} (1993) 4337,
  [hep-ph/9203216];
%
 J.~M.~Cornwall, D.~N.~Levin and G.~Tiktopoulos,
 {\it Derivation Of Gauge Invariance From High-Energy Unitarity Bounds On The S -
   Matrix,}
  Phys.\ Rev.\ D {\bf 10}, 1145 (1974)
  [Erratum-ibid.\ D {\bf 11}, 972 (1975)].
  %


\bibitem{SoftBreaking}
  L.~Girardello and M.~T.~Grisaru,
  {\it Soft Breaking Of Supersymmetry,}
  Nucl.\ Phys.\ B {\bf 194}, 65 (1982);
%
  S.~K.~Soni and H.~A.~Weldon,
  {\it Analysis Of The Supersymmetry Breaking Induced By N=1 Supergravity
  Theories,}
  Phys.\ Lett.\ B {\bf 126}, 215 (1983);
%
   L.~J.~Hall, J.~D.~Lykken and S.~Weinberg,
  {\it Supergravity As The Messenger Of Supersymmetry Breaking,}
  Phys.\ Rev.\ D {\bf 27}, 2359 (1983);
%
  D.~J.~H.~Chung, L.~L.~Everett, G.~L.~Kane, S.~F.~King, J.~D.~Lykken and L.~T.~Wang,
  {\it The soft supersymmetry-breaking Lagrangian: Theory and applications,}
  Phys.\ Rept.\  {\bf 407}, 1 (2005),
  [hep-ph/0312378].
%
\bibitem{BI}
%
  A.~Brignole, L.~E.~Ibanez and C.~Munoz,
  {\it Towards a theory of soft terms for the
    supersymmetric Standard Model,}
  Nucl.\ Phys.\ B {\bf 422} (1994) 125
  [Erratum-ibid.\ B {\bf 436} (1995) 747],
  [hep-ph/9308271];
%
  V.~S.~Kaplunovsky and J.~Louis,
  {\it Model independent analysis of soft terms
   in effective supergravity and in string theory,}
  Phys.\ Lett.\ B {\bf 306}, 269 (1993),
  [hep-th/9303040].

\bibitem{GVW}
S. Gukov, C. Vafa and E. Witten, 
Nucl. Phys. {\bf B584}, 69 (2000).

\bibitem{BHK}
  M.~Berg, M.~Haack and B.~Kors,
  {\it Loop corrections to volume moduli and inflation
   in string theory,}
  Phys.\ Rev.\ D {\bf 71}, 026005 (2005),
  [hep-th/0404087].

\bibitem{hepth0607050}
D.~Baumann, A.~Dymarsky, I.~R.~Klebanov, J.~Maldacena,
  L.~McAllister and A.~Murugan,
   {\it On D3-brane potentials in compactifications
    with fluxes and wrapped D-branes,}
  [hep-th/0607050].

\bibitem{hepth0209200}
M.~Grana,
  {\it MSSM parameters from supergravity backgrounds,}
  Phys.\ Rev.\ D {\bf 67} (2003) 066006,
  [hep-th/0209200].

\bibitem{hepth0311241}
  P.~G.~Camara, L.~E.~Ibanez and A.~M.~Uranga,
  {\it Flux-induced SUSY-breaking soft terms,}
  Nucl.\ Phys.\ B {\bf 689} (2004) 195,
  [hep-th/0311241].

\bibitem{dellis}
  G.~A.~Diamandis, J.~R.~Ellis, A.~B.~Lahanas and D.~V.~Nanopoulos,
  {\it Vanishing scalar masses in no scale supergravity,}
  Phys.\ Lett.\ B {\bf 173}, 303 (1986).

\bibitem{hepth0408036}
  P.~G.~Camara, L.~E.~Ibanez and A.~M.~Uranga,
  {\it Flux-induced SUSY-breaking soft terms on D7-D3 brane systems,}
  Nucl.\ Phys.\ B {\bf 708}, 268 (2005),
  [hep-th/0408036].

\bibitem{GP}
  M.~Grana and J.~Polchinski,
  {\it Supersymmetric three-form flux perturbations on AdS(5),}
  Phys.\ Rev.\ D {\bf 63} (2001) 026001,
  [hep-th/0009211];
  %
  M.~Grana and J.~Polchinski,
  {\it Gauge / gravity duals with holomorphic dilaton,}
  Phys.\ Rev.\ D {\bf 65} (2002) 126005,
  [hep-th/0106014].

\bibitem{QuevUrang}
J.~F.~G.~Cascales, M.~P.~Garcia del Moral, F.~Quevedo and A.~M.~Uranga,
   {\it Realistic D-brane models on warped throats: Fluxes, hierarchies and  moduli
  stabilization,}
  JHEP {\bf 0402} (2004) 031,
  [hep-th/0312051].

\bibitem{hepph9812397}
L.~E.~Ib\'a\~nez, C.~Munoz and S.~Rigolin,
{\it Aspects of type I string phenomenology,}
Nucl.\ Phys.\ B {\bf 553}, 43 (1999),
[hep-ph/9812397].

\bibitem{hepth9907189}
  D.~S.~Freed and E.~Witten,
  {\it Anomalies in string theory with D-branes,}
  [hep-th/9907189].

\bibitem{hepth0409098}
H.~Jockers and J.~Louis,
{\it The effective action of D7-branes in N = 1 Calabi-Yau orientifolds,}
Nucl.\ Phys.\ B {\bf 705}, 167 (2005),
[hep-th/0409098].


\bibitem{hepth0501139}
D.~Lust, P.~Mayr, S.~Reffert and S.~Stieberger,
{\it F-theory flux, destabilization of orientifolds and soft terms on
D7-branes,}
[hep-th/0501139].

\bibitem{hepth0406092}
D.~Lust, S.~Reffert and S.~Stieberger,
 ``Flux-induced soft supersymmetry breaking in chiral type IIb  orientifolds
  with D3/D7-branes,''
  Nucl.\ Phys.\ B {\bf 706}, 3 (2005)
  [arXiv:hep-th/0406092].

\bibitem{hepth0410074}
  D.~Lust, S.~Reffert and S.~Stieberger,
  ``MSSM with soft SUSY breaking terms from D7-branes with fluxes,''
  Nucl.\ Phys.\ B {\bf 727}, 264 (2005)
  [arXiv:hep-th/0410074].

\bibitem{nilles}
K.~Choi, A.~Falkowski, H.~P.~Nilles and M.~Olechowski,
  {\it Soft supersymmetry breaking in KKLT flux compactification,}
  Nucl.\ Phys.\ B {\bf 718} (2005) 113,
  [hep-th/0503216].

\bibitem{largev}
  J.~P.~Conlon, F.~Quevedo and K.~Suruliz,
  {\it Large-volume flux compactifications: Moduli spectrum and D3/D7 soft
  supersymmetry breaking,}
  JHEP {\bf 0508} (2005) 007,
  [hep-th/0505076];
%
  B.~C.~Allanach, F.~Quevedo and K.~Suruliz,
  {\it Low-energy supersymmetry breaking from string flux compactifications:
  Benchmark scenarios,}
  JHEP {\bf 0604}, 040 (2006),
  [hep-ph/0512081];
%
 J.~P.~Conlon and F.~Quevedo,
  {\it Gaugino and scalar masses in the landscape,}
  JHEP {\bf 0606}, 029 (2006),
  [hep-th/0605141];
J.~P.~Conlon, S.~S.~Abdussalam, F.~Quevedo and K.~Suruliz,
  {\it Soft SUSY Breaking Terms for Chiral Matter in IIB String
  Compactifications,}
  [hep-th/0610129].


\bibitem{botomup}
G.~Aldazabal, L.~E.~Ibanez, F.~Quevedo and A.~M.~Uranga,
  {\it D-branes at singularities: A bottom-up approach to the string  embedding of
  the standard model,}
  JHEP {\bf 0008} (2000) 002,
  [hep-th/0005067];
D.~Berenstein, V.~Jejjala and R.~G.~Leigh,
  {\it The standard model on a D-brane,}
  Phys.\ Rev.\ Lett.\  {\bf 88} (2002) 071602,
  [hep-ph/0105042];
L.~F.~Alday and G.~Aldazabal,
  {\it In quest of 'just' the standard model on D-branes at a singularity,}
  JHEP {\bf 0205} (2002) 022,
  [hep-th/0203129];
H.~Verlinde and M.~Wijnholt,
  {\it Building the standard model on a D3-brane,}
  [hep-th/0508089].

\bibitem{EWDSSB}
  E.~Witten,
  ``Dynamical Breaking Of Supersymmetry,''
  Nucl.\ Phys.\  B {\bf 188} (1981) 513.

\bibitem{IntScale}
K.~Benakli, {\it Phenomenology of low quantum gravity scale
models,} Phys.\ Rev.\ D {\bf 60}, 104002 (1999), [hep-ph/9809582];
C.~P.~Burgess, L.~E.~Ibanez and F.~Quevedo, {\it Strings at the
intermediate scale or is the Fermi scale dual to the  Planck
scale?} Phys.\ Lett.\ B {\bf 447}, 257 (1999), [hep-ph/9810535].

\bibitem{bbcq}
V.~Balasubramanian, P.~Berglund, J.~P.~Conlon and F.~Quevedo,
  {\it Systematics of moduli stabilisation in Calabi-Yau flux
  compactifications,}
  JHEP {\bf 0503} (2005) 007,
  [hep-th/0502058].

\bibitem{DD}
F.~Denef and M.~R.~Douglas,
  {\it Distributions of flux vacua,}
  JHEP {\bf 0405}, 072 (2004),
  [hep-th/0404116].

\bibitem{MH}
A.~Hebecker and J.~March-Russell,
  {\it The ubiquitous throat,}
  [hep-th/0607120].


\end{thebibliography}
\end{document}